\documentclass[useAMS,usenatbib]{mn2e}
\usepackage{amssymb}
\usepackage{amsmath}
\usepackage{graphicx}
\usepackage[draft]{hyperref} 
\usepackage{color}
\usepackage{soul}
\usepackage[normalem]{ulem}
\usepackage[utf8]{inputenc}
\newcommand{\nodata}{}
\begin{document}

\title[BAO in the correlation function of LOWZ \& CMASS]{The clustering of galaxies in the SDSS-III Baryon Oscillation Spectroscopic Survey: Baryon Acoustic Oscillations in the correlation function of LOWZ and CMASS galaxies in Data Release 12}

\author[A. J. Cuesta et al.]{\parbox{\textwidth}{
   Antonio J.~Cuesta$^1$\thanks{Email: ajcuesta@icc.ub.edu},
   Mariana Vargas-Maga{\~n}a$^{2,3,4}$,
   Florian Beutler$^{5,6}$,
   Adam S. Bolton$^{7}$,
   Joel R. Brownstein$^{7}$,
   Daniel J. Eisenstein$^{8}$,
   H{\'e}ctor Gil-Mar{\'\i}n$^{9}$,
   Shirley Ho$^{3,4}$,
   Cameron K. McBride$^{8}$,
   Claudia Maraston$^{9}$,
   Nikhil Padmanabhan$^{10}$,
   Will J. Percival$^{9}$,
   Beth A. Reid$^{5,6}$,
   Ashley J. Ross$^{11}$,
   Nicholas P. Ross$^{12,13}$,
   Ariel G. S\'anchez$^{14}$,
   David J. Schlegel$^{5,6}$,
   Donald P. Schneider$^{15,16}$,
   Daniel Thomas$^{9}$,
   Jeremy Tinker$^{17}$,
   Rita Tojeiro$^{18}$,
   Licia Verde$^{1,19,20,21}$
   and Martin White$^{5,6,22,23}$
 }
\vspace{3mm}
\\
$^{1}$Institut de Ci{\`e}ncies del Cosmos (ICCUB), Universitat de Barcelona (IEEC-UB), Mart{\'\i} i Franqu{\`e}s 1, E08028 Barcelona, Spain \\
$^{2}$Instituto de F{\'\i}sica, UNAM, P.O. Box 20-364, 01000 M\'exico D.F., Mexico \\
$^{3}$Bruce and Astrid McWilliams Center for Cosmology, Department of Physics, Carnegie Mellon University, 5000 Forbes Ave.,\\
    Pittsburgh, PA 15213, USA \\
$^{4}$Department of Physics, Carnegie Mellon University, 5000 Forbes Ave., Pittsburgh, PA 15217 \\
$^{5}$Berkeley Center for Cosmological Physics, Department of Physics, University of California, Berkeley, CA, 94720, USA \\
$^{6}$Lawrence Berkeley National Lab, 1 Cyclotron Rd, Berkeley, CA 94720, USA \\
$^{7}$Department of Physics and Astronomy, University of Utah, 115 S 1400 E, Salt Lake City, UT 84112, USA \\
$^{8}$Harvard University, Harvard-Smithsonian Center for Astrophysics, 60 Garden St., Cambridge, MA 02138, USA \\
$^{9}$Institute of Cosmology \& Gravitation, Dennis Sciama Building, University of Portsmouth, Portsmouth, PO1 3FX, UK \\
$^{10}$Department of Physics, Yale University, 260 Whitney Avenue, New Haven, CT 06520, USA \\
$^{11}$Center for Cosmology and AstroParticle Physics, The Ohio State University, Columbus, OH 43210, USA \\
$^{12}$Institute for Astronomy, University of Edinburgh,  Royal Observatory, Blackford Hill Edinburgh EH9 3HJ, United Kingdom \\
$^{13}$Department of Physics, Drexel University, 3141 Chestnut Street, Philadelphia, PA 19104, USA \\
$^{14}$Max-Planck-Institut f{\"u}r Extraterrestrische Physik, Giessenbachstrasse 1, D-85748 Garching, Germany \\
$^{15}$Department of Astronomy and Astrophysics, The Pennsylvania State University, University Park, PA 16802 \\
$^{16}$Institute for Gravitation and the Cosmos, The Pennsylvania State University, University Park, PA 16802 \\
$^{17}$Center for Cosmology and Particle Physics, Department of Physics, New York University \\
$^{18}$School of Physics and Astronomy, University of St Andrews, North Haugh, St Andrews KY16 9SS, UK \\
$^{19}$ICREA (Instituci\'o Catalana de Recerca i Estudis Avan\c{c}ats) \\
$^{20}$Radcliffe Institute for Advanced Study, Harvard University, MA 02138, USA \\
$^{21}$Institute of Theoretical Astrophysics, University of Oslo, 0315 Oslo, Norway \\
$^{22}$Department of Astronomy, University of California, Berkeley, CA 94720, USA \\
$^{23}$Department of Physics, University of California, Berkeley, CA 94720, USA \\
}

\maketitle

\begin{abstract}
We present distance scale measurements from the baryon acoustic oscillation signal in the CMASS and LOWZ samples from the Data Release 12 of the Baryon Oscillation Spectroscopic Survey (BOSS). The total volume probed is 14.5 Gpc$^3$, a 10 per cent increment from Data Release 11. From an analysis of the spherically averaged correlation function, we infer a distance to $z=0.57$ of $D_V(z)r^{\rm fid}_{\rm d}/r_{\rm d}=2028\pm21$ Mpc and a distance to $z=0.32$ of $D_V(z)r^{\rm fid}_{\rm d}/r_{\rm d}=1264\pm22$ Mpc assuming a cosmology in which $r^{\rm fid}_{\rm d}=147.10$ Mpc. From the anisotropic analysis, we find an angular diameter distance to $z=0.57$ of $D_{\rm A}(z)r^{\rm fid}_{\rm d}/r_{\rm d}=1401\pm21$ Mpc and a distance to $z=0.32$ of $981\pm20$ Mpc, a 1.5 per cent and 2.0 per cent measurement respectively. The Hubble parameter at $z=0.57$ is $H(z)r_{\rm d}/r^{\rm fid}_{\rm d}=100.3\pm3.7$ km s$^{-1}$ Mpc$^{-1}$ and its value at $z=0.32$ is $79.2\pm5.6$ km s$^{-1}$ Mpc$^{-1}$, a 3.7 per cent and 7.1 per cent measurement respectively. These cosmic distance scale constraints are in excellent agreement with a $\Lambda$CDM model with cosmological parameters released by the recent \textit{Planck} 2015 results. 
\end{abstract}

\begin{keywords}
  cosmology: observations, distance scale, large-scale structure of Universe
\end{keywords}

\section{Introduction}
The Data Release 12 (DR12, \citealt{DR12})\footnote{http://sdss.org/dr12} of the Baryon Oscillation Spectroscopic Survey (BOSS, \citealt{BOSS}) represents a major milestone in the history of baryon acoustic oscillation (BAO) observations, and in general, of cosmic distance scale measurements. The unprecedented precision goal of 1 per cent in a cosmological distance was achieved in Data Release 11 \citep{Anderson2014} and has not been matched since then, even by local expansion rate measurements. Improvements are expected in the next few years extending to higher redshifts with the extended BOSS survey \citep{eBOSS} and HETDEX \citep{HETDEX}. Substantial improvements are not expected until results are available from the next generation of experiments, including EUCLID \citep{EuclidAssesment, EuclidDefinition}, LSST \citep{LSST}, SKA \citep{SKA}, WFIRST \citep{WFIRST, WFIRSTReport}, and DESI \citep{DESIOverview, DesiWhitePaper}.

This breakthrough is the continuation of a ten-year history of BAO observations. Early BAO measurements by \cite{EisensteinBAO}, using a previous incarnation of the SDSS survey, and \cite{ColeBAO2005}, using the 2dF survey, paved the way for modern BAO measurements from galaxy surveys such as the 6dFGS \citep{Beutler2011} and WiggleZ \citep{Blake2011}. Later measurements reconstructed the linear density field in order to improve distance scale constraints \citep{Reconstruction,Padmanabhan2012}, including existing distance measurements \citep{Padmanabhan2012,Kazin2014,MGS} and the new BOSS measurements \citep{DR9,Anderson2014}.

BOSS has populated the distance-redshift diagram with four new data points, two from the clustering of galaxies \citep{Tojeiro2014,Anderson2014} and two from the two-point function of the transmission flux in the Lyman-$\alpha$ forest \citep{Font2014,Delubac2014}. The enormous volume probed by these samples has been key to providing low-uncertainty distance scale measurements, which have become an invaluable input for most state-of-the art cosmological analyses.

Baryon acoustic oscillations distance measurements using BOSS galaxies have traditionally been determined using two different galaxy samples: The ``Constant Stellar Mass'' sample, or CMASS, covering redshifts in the range $0.43<z<0.70$ and a fiducial redshift of 0.57, and the low-redshift sample, or LOWZ, covering redshifts of $0.15<z<0.43$ with an effective redshift of 0.32 \citep[][companion paper]{Reid2015}.

The Data Release 12 represents an increment of 10 per cent in area, volume, and number of galaxies over Data Release 11. The main paper \citet[in preparation]{Anderson2015} (hereafter the \textit{Final Data Release} paper) offers an analysis of the same DR12 sample not split into LOWZ and CMASS, but combined optimally together. Here instead as part of the BOSS legacy, we present the distance scale measurements from the traditional LOWZ and CMASS measurements, which serve a two-fold purpose. The constraining power of the results in \citet[in preparation]{Anderson2015} can be tested against the traditional analysis (this paper), hence being a benchmark of the new analysis techniques. Moreover, the results presented here can be readily compared with previous BOSS Data Releases, which provides more transparency to our final results.

This analysis not only benefits from new data. We also take advantage from an updated version of the systematic weights \citep[in preparation]{Ross2015} to account for spurious large scale fluctuations in the galaxy number density due to observational systematic effects. We have also updated the set of mocks to compute the covariance matrix to the new Quick-Particle-Mesh, or QPM, mocks \citep{QPM}. The QPM mocks, which are generated using a new methodology and with a cosmology closer to current constraints than the mocks used in DR11, result in an improved measurement not only in terms of the formal statistical errors and covariances but also offer a reliable, more robust determination of the cosmic distance scale and its uncertainty.

This paper is organized as follows: in Section~\ref{sec:data} we describe the final statistics of the LOWZ and CMASS samples of BOSS. Section~\ref{sec:methods} discusses the mocks and tests our fitting techniques. Section~\ref{sec:fitting} presents the results of the isotropic and anisotropic fittings of the two-point function of LOWZ and CMASS in configuration space (and compares them with those in Fourier space), Section~\ref{sec:cosmo} discusses the cosmological implications of these distance measurements, and Section~\ref{sec:conclusion} summarizes our conclusions.

\section{Datasets}
\label{sec:data}
This paper uses data from the Data Release 12 \citep{DR12} of the Baryon Oscillation Spectroscopic Survey (BOSS) \citep{SDSS3Technical}. The BOSS survey uses the SDSS 2.5 meter telescope at Apache Point Observatory \citep{SDSSTelescope} and the spectra are obtained using the double-armed BOSS spectrograph \citep{BOSSSpectrograph}. The data are then reduced using the algorithms described in \cite{BOSSPipeline}. 

The target selection of the CMASS and LOWZ samples, together with the algorithms used to create large scale structure catalogues (the \textsc{mksample} code), are presented in the companion paper \cite{Reid2015}. 

The LOWZ sample contains 361,762 galaxies in the range $0.15<z<0.43$, with 248,237 in the North Galactic Cap (NGC) and 113,525 in the South Galactic Cap (SGC). The CMASS sample contains 777,202 galaxies in the range $0.43<z<0.70$, with 568,776 in the NGC and 208,426 in the SGC. This total of 1,138,964 galaxies is used in our analysis (see Table~\ref{tab:stats}).

\begin{table}
\centering
\caption{Statistics of the LOWZ and CMASS galaxy samples used in this paper.}
\begin{tabular}{|l|rrr|}
\hline
\hline
            & NGC     & SGC     & Total     \\
\hline
LOWZ        & 248,237 & 113,525 &   361,762 \\
CMASS       & 568,776 & 208,426 &   777,202 \\
LOWZ+CMASS  & 817,013 & 321,951 & 1,138,964 \\
\hline
\end{tabular}
\label{tab:stats}
\end{table}

The area covered by these samples is shown in Table~\ref{tab:area}, including a comparison with the coverage in Data Release 11. The sky coverage in CMASS sample has increased by 11.9 per cent whereas that of the LOWZ sample has increased by 13.6 per cent. The LOWZ area is slightly smaller than the CMASS area mainly because some regions of LOWZ were targeted with a different selection \citep[][companion paper]{Reid2015}. Those regions however will be included in the analysis shown in \citet[in preparation]{Anderson2015}. 

\begin{table}
\centering
\caption{Sky coverage in LOWZ and CMASS samples (effective area, in deg$^2$).}
\begin{tabular}{|l|rrr|}
\hline
\hline
            & NGC     & SGC     & Total     \\
\hline
LOWZ DR11       & 5290.82 & 2050.60 & 7341.42 \\
LOWZ DR12       & 5836.21 & 2501.26 & 8337.47 \\
\hline
CMASS DR11      & 6307.94 & 2068.96 & 8376.90 \\
CMASS DR12      & 6851.42 & 2524.67 & 9376.09 \\
\hline
\end{tabular}
\label{tab:area}
\end{table}

The total volume (assuming our fiducial cosmology described in Section~\ref{sec:methods}) that the LOWZ and CMASS galaxies occupy amounts to a total of 14.5 Gpc$^3$, out of which 10.8 Gpc$^3$ corresponds to CMASS and 3.7 Gpc$^3$ corresponds to LOWZ. We also compute the effective volume $V_{\rm eff}$, defined as:
\begin{equation}
V_{\rm eff} = \int{\mathrm{d}V\left(\frac{n(z)P_0}{1+n(z)P_0}\right)^2}
\end{equation}
where $P_0$ is an estimate of the amplitude of the power spectrum at the BAO scale, here assumed to be $P_0$=20,000 $h^{-3}$Mpc$^3$ (as in \citealt{Anderson2014}), and $n(z)$ is the galaxy number density. In \citet[in preparation]{Anderson2015} the value used is 10,000 $h^{-3}$Mpc$^3$ following \cite{FontDESI2014}. For a comparison of volumes in different cosmologies and $P_0$ values see Table~\ref{tab:volume}.

\begin{table}
\centering
\caption{Effective volume (in Gpc$^3$) of the LOWZ and CMASS samples for different values of the matter density $\Omega_m$, the Hubble parameter $h$, and the amplitude of the matter power spectrum at the BAO scale $P_0$.}
\begin{tabular}{|l|rrr|}
\hline
\hline
            & LOWZ     & CMASS     & Total     \\
\hline
$P_0=$20,000 $h^{-3}$Mpc$^3$ & & & \\
$\Omega_m$=0.274, $h=0.700$ & 2.65 & 6.65 & 9.30 \\
$\Omega_m$=0.290, $h=0.700$ & 2.62 & 6.55 & 9.17 \\
$\Omega_m$=0.310, $h=0.676$ & 2.87 & 7.14 &10.01 \\
\hline
$P_0=$10,000 $h^{-3}$Mpc$^3$ & & & \\
$\Omega_m$=0.274, $h=0.700$ & 2.00 & 4.70 & 6.70 \\
$\Omega_m$=0.290, $h=0.700$ & 1.98 & 4.65 & 6.63 \\
$\Omega_m$=0.310, $h=0.676$ & 2.18 & 5.11 & 7.29 \\
\hline
\end{tabular}
\label{tab:volume}
\end{table}

\begin{figure*}
\includegraphics[width=\textwidth]{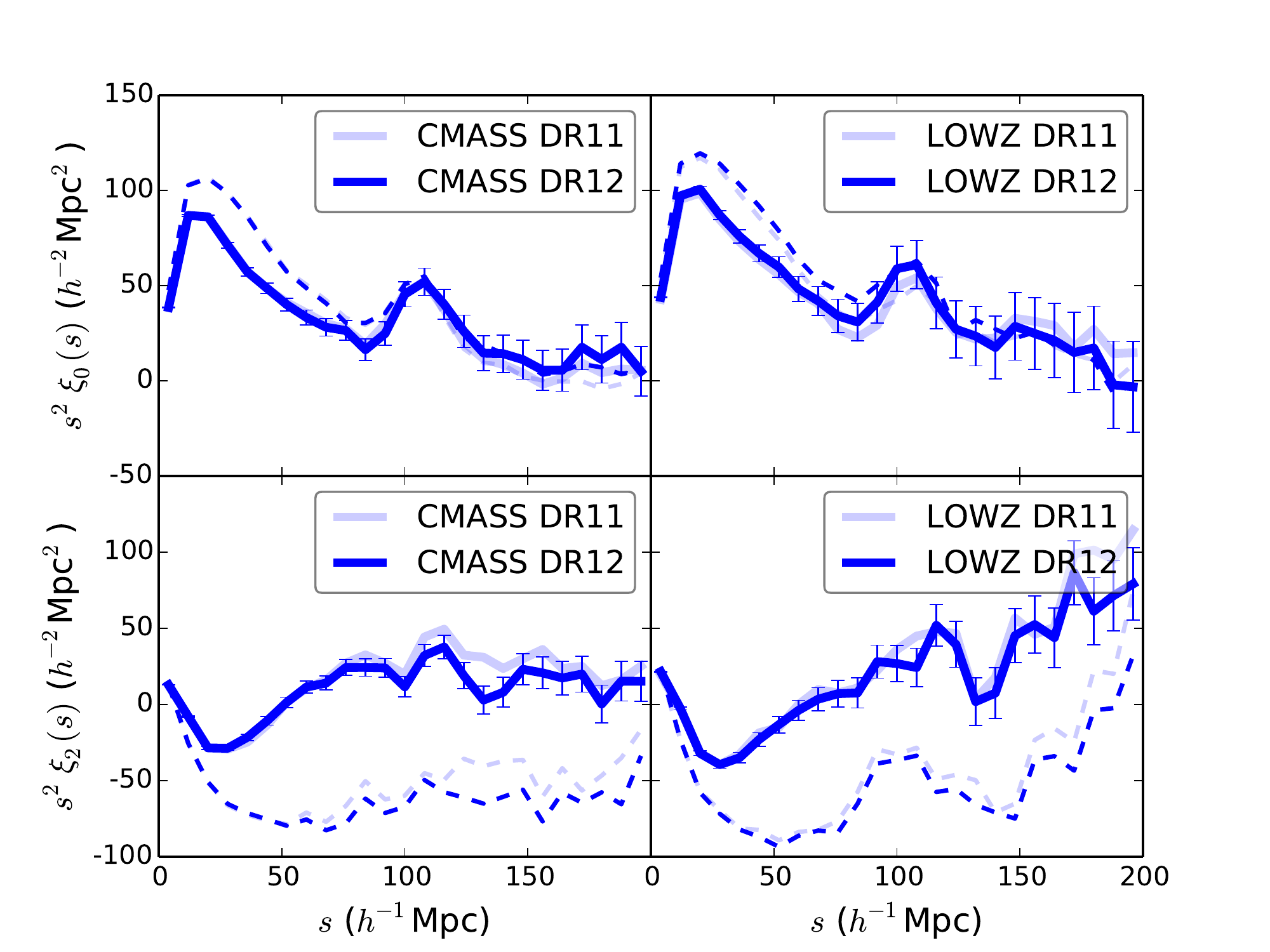}
\caption{Monopole (top) and quadrupole (bottom) of the CMASS and LOWZ correlation functions assuming our fiducial cosmology. Left panels show the CMASS correlation function, whereas right panels present the LOWZ correlation function. In all panels the dashed line indicates the correlation function pre-reconstruction. The lighter shade is the DR11 version for comparison. Error bars represent the square root of the diagonal elements of the covariance matrix.}
\label{fig:dr12}
\end{figure*}

We then compute the FKP-weighted \citep{FKP} correlation functions using the Landy-Szalay estimator \citep{Landy1993}, with a random catalogue\footnote{The size of the random catalogue is 50 times the size of the data samples, and in the case of the QPM mocks we use 20 times the size of the mock catalogues.} generated by the \textsc{mksample} code \citep[see][companion paper]{Reid2015} to match the geometry, redshift distribution, and completeness of the survey. These functions include the corrections for systematic effects described in \citet[in preparation]{Ross2015}, which account for correlations between the observed galaxy density of the CMASS sample and stellar density in the sky and seeing. We also include weights that correct for close pairs (fibre collisions) and redshift failures in these samples. A detailed description of the observational systematic weights and their effect on the measured clustering will be provided in \citet[in preparation]{Ross2015}. As shown in \cite{Ross2012}, these weights compensate for the systematic effect by interpolating the observed deficit in the number density of galaxies as a function of the systematic, and weighting by the inverse of this deficit. The systematic that has a larger effect in the measured correlation function is stellar density, with seeing providing a more modest correction. Fibre collision weights are very significant at small scales, although their contribution to the clustering is negligible at the BAO scale. Redshift failure weights also show a rather small effect in the measured clustering. Again, this topic will be revisited in the context of the DR12 samples in \citet[in preparation]{Ross2015}, where any impact on the measured BAO scale is found to be negligible. The resulting correlation functions for CMASS and LOWZ are shown in Figure~\ref{fig:dr12}, where we display the pre-reconstruction correlation functions with a dashed line. As in \cite{Anderson2014}, we also apply density field reconstruction \citep{Padmanabhan2012} to our samples, which we test on mock galaxy samples in Section~\ref{sec:methods}. The resulting post-reconstruction correlation functions are plotted with a solid line. Also displayed for reference is the correlation function from the previous Data Release 11 (re-computed using the fiducial cosmology used in this paper) with a fainter line.

\section{Methodology}
\label{sec:methods}

The mock catalogues and error estimates are computed with a fiducial cosmology that is close to the best-fit Planck+BOSS cosmology, such that they faithfully produce the covariances and fitting errors of the data. Our fiducial cosmology is given by the following set of cosmological parameters\footnote{This is a slightly different choice from that of the Final Data Release paper \citep[in preparation]{Anderson2015} in which $\Omega_m$=0.31, $\Omega_{\Lambda}$=0.69, $\Omega_k$=0, $\Omega_bh^2$=0.022, $\Omega_{\nu}h^2=0.00064$, $w=-1$, $w_a=0$, $h=0.676$, $n_s=0.97$, and $\sigma_8$=0.8}: $\Omega_m$=0.29, $\Omega_{\Lambda}$=0.71, $\Omega_k$=0, $\Omega_bh^2$=0.02247, $\Omega_{\nu}h^2=0.0$, $w=-1$, $w_a=0$, $h=0.7$, $n_s=0.97$, and $\sigma_8$=0.8. The choice of this cosmology is motivated by our Planck+BAO (i.e. Planck+LOWZ+CMASS+6dF+LyA) constraints in \cite{Anderson2014} in the $\Lambda$CDM model, and the fact that the fiducial cosmology in \cite{Anderson2014} corresponds to a value of $\Omega_mh^3$ that lies more than 6$\sigma$ away from the tight constraints from Planck Cosmic Microwave Background data \citep{Planck2015Cosmo}. In this fiducial cosmology\footnote{For comparison, in the fiducial cosmology of \citet[in preparation]{Anderson2015} the volume-averaged distance to redshift $z=0.32$ is $D_V(0.32)=1271.2215$Mpc, $D_A(0.32)=991.9952$ Mpc, $H(0.32)=80.07077$ km s$^{-1}$ Mpc$^{-1}$, the distance to redshift $z=0.57$ is $D_V(0.57)=2059.5562$ Mpc, $D_A(0.57)=1388.298$ Mpc, $H(0.57)=92.92644$ km s$^{-1}$ Mpc$^{-1}$ and the sound horizon scale is $r_{\rm{d,fid}}=147.781$ Mpc.}, the volume-averaged distance to redshift $z=0.32$ is $D_V(0.32)=1235.28$ Mpc, $D_A(0.32)=962.43$ Mpc, $H(0.32)=82.142$ km s$^{-1}$ Mpc$^{-1}$, the distance to redshift $z=0.57$ is $D_V(0.57)=2009.55$ Mpc, $D_A(0.57)=1351.13$ Mpc, $H(0.57)=94.753$ km s$^{-1}$ Mpc$^{-1}$ and the sound horizon scale is $r_{\rm{d,fid}}=147.10$ Mpc. The sound horizon is evaluated using \textsc{CAMB} \citep{Lewis2000}\footnote{http://camb.info}.

The fiducial value of the sound horizon scale used in the analysis in \cite{Anderson2014} ($r_{\rm{d,fid}}=$149.28 Mpc) is now ruled out by more than $6\sigma$ from its inferred value using Planck temperature and polarization data in a $\Lambda$CDM model where the effective number of relativistic species is set to the standard value of $N_{\rm eff}=$3.046. The fiducial value used in \citet[in preparation]{Anderson2015} is $r_{\rm{d,fid}}=$147.78 Mpc, consistent with the constraints from the Planck observations for a standard $\Lambda$CDM model.

The analysis of the clustering of galaxies in Data Release 12 uses two sets of mocks in order to estimate the covariance matrix. These are the QPM mocks \citep{QPM} and the MultiDark-Patchy BOSS DR12 mocks, hereafter MD-Patchy\footnote{http://data.sdss3.org/datamodel/index-files.html} (\citealt[][-companion paper-]{PATCHY}, \citealt{PATCHYtheory}). Both were generated to match the footprint and number density of the CMASS and LOWZ samples from Data Release 12. We have fitted the DR12 correlation function using both covariance matrices; the QPM mocks return a slightly larger error bar. Since we find no compelling reason to choose one set of mocks over the other, we will adopt a conservative approach and from now on we discuss the results that quote a larger statistical uncertainty, corresponding to those using the QPM covariance matrix only. The values of cosmological parameters for both cosmologies are shown in Table~\ref{tab:fidcosmo} and the fiducial distances to $z=0.32$ and $z=0.57$ in Table~\ref{tab:fiddist}. 

\begin{table*}
\centering
\caption{Cosmological parameters of the QPM and MD-Patchy mock catalogues. Our choice for the fiducial cosmology corresponds to the cosmology of the QPM mocks. For comparison we include the values corresponding to the fiducial cosmology in Anderson et al. (2016).}
\begin{tabular}{lccccccc}
\hline
\hline
                    & $h$ & $\Omega_b h^2$ & $\Omega_m$ & $\Omega_{\Lambda}$ & $n_s$ & $\sigma_8$ \\
\hline
QPM \& this work    & 0.7     & 0.02247 & 0.29     & 0.71     & 0.97   & 0.8 \\
MD-Patchy           & 0.6777  & 0.02214 & 0.307115 & 0.692885 & 0.9611 & 0.8288 \\
\cite{Anderson2015} & 0.676   & 0.022   & 0.31     & 0.69     & 0.97   & 0.8  \\
\hline
\end{tabular}
\label{tab:fidcosmo}
\end{table*}

\begin{table*}
\centering
\caption{Fiducial distances and Hubble parameters for the cosmologies of the QPM and MD-Patchy mock catalogues, computed at the fiducial redshifts of LOWZ ($z=0.32$) and CMASS ($z=0.57$) assuming a $\Lambda$CDM cosmological model. Our choice for the fiducial cosmology corresponds to the cosmology of the QPM mocks. The sound horizon at radiation drag $r_{\rm d}$ was evaluated using \textsc{CAMB}. For comparison we include the values corresponding to the fiducial cosmology in Anderson et al. (2015).}
\begin{tabular}{lccccccc}
\hline
\hline
                    & $r_{\rm d}$ & $D_{\rm A}(z=0.32)$ & $H(z=0.32)$ & $D_V(z=0.32)$ & $D_{\rm A}(z=0.57)$ & $H(z=0.57)$ & $D_V(z=0.57)$ \\
                    &  (Mpc)      & (Mpc)               & (km s$^{-1}$ Mpc$^{-1}$) & (Mpc) & (Mpc)               & (km s$^{-1}$ Mpc$^{-1}$) & (Mpc) \\
\hline
QPM \& this work    & 147.10 & 962.43 & 82.142 & 1235.28 & 1351.13 & 94.753 & 2009.55 \\
MD-Patchy           & 147.66 & 990.16 & 80.165 & 1269.16 & 1386.35 & 92.956 & 2057.41 \\
\cite{Anderson2015} & 147.78 & 992.00 & 80.071 & 1271.22 & 1388.30 & 92.926 & 2059.56 \\
\hline
\end{tabular}
\label{tab:fiddist}
\end{table*}

The data catalogues have been carefully tested for systematics. A thorough study of systematics in LOWZ and CMASS galaxy samples is presented in \citet[in preparation]{Ross2015}. All the systematic effects found to correlate with galaxy density of these samples are compensated by assigning weights to each galaxy. After including all these systematic weights, and combining them with the corrections from close pairs and redshift failures, we compute the clustering of the resulting samples in our fiducial cosmology. We adopt the reconstruction technique \citep{Reconstruction} which has been applied regularly in galaxy surveys since \cite{Padmanabhan2012} to partially remove the effect of non-linearities on the uncertainties in cosmic distance measurements \citep{NonLinearBAO2,NonLinearBAO}. A bias parameter of $b=1.85$ for both samples \citep{Anderson2014,Tojeiro2014} and a redshift space distortion parameter $\beta=b^{-1} \mathrm{d}\ln D /\mathrm{d} \ln a$ from the QPM cosmology ($\beta=0.4128$ for the central redshift of CMASS, $\beta=0.3628$ for LOWZ) is assumed. For the mocks, a bias of $b=2.1$ and a value of $\beta=0.3607$ for the central redshift of CMASS, $\beta=0.3353$ for LOWZ are adopted. The difference in the bias is due to the relative amplitude of the clustering of the mocks and the data at scales of 30 $h^{-1}$Mpc with respect to the clustering of the dark matter in our fiducial cosmology. In all cases a Gaussian kernel of 15 $h^{-1}$Mpc is applied to smooth the galaxy density field when applying reconstruction. The effects of the particular choice of the smoothing length are studied in detail in the companion paper \cite{Vargas-Magana2015a}.

\begin{figure*}
\centering
\includegraphics[width=\textwidth]{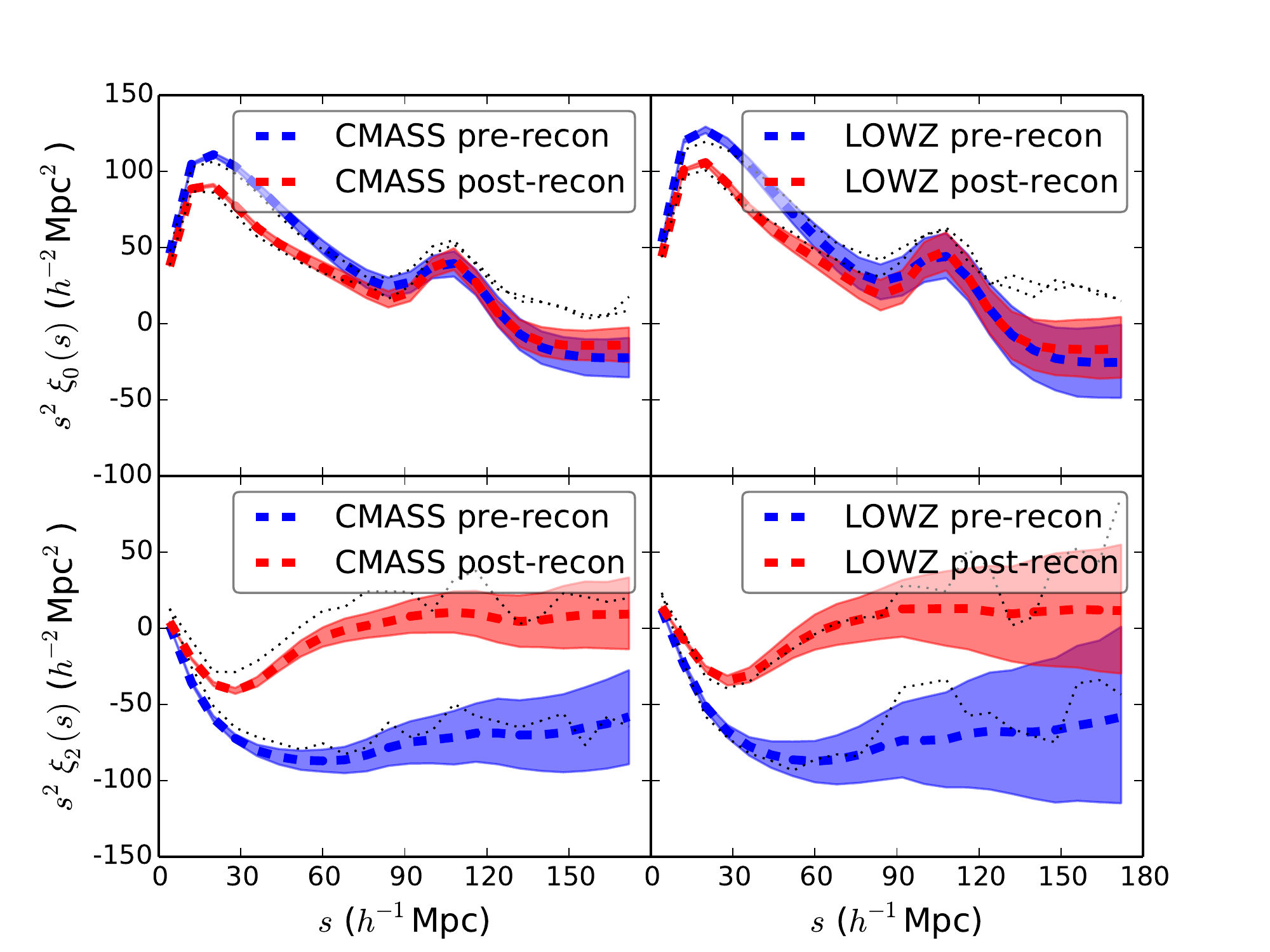}
\caption{The correlation function of QPM mocks, pre-reconstruction (blue) and post-reconstruction (red). The upper panels show the average and standard deviation (dashed line and shaded regions) of the monopole of the CMASS (left) and LOWZ (right) correlation function of the QPM mocks. Bottom panels display the average and standard deviation of the quadrupole of the QPM correlation functions for CMASS (left) and LOWZ (right). A dotted line shows the clustering of the data for comparison.}
\label{fig:qpm}
\end{figure*}

Figure~\ref{fig:qpm} shows the monopole and quadrupole of the QPM mock correlation functions pre-reconstruction (blue) and post-reconstruction (red) for the CMASS and LOWZ samples. The shaded region represents the standard deviation of the mock correlation functions around their average, which is displayed with a dashed line. For reference we include the observed correlation functions from Data Release 12 as dotted lines. We note that as in \cite{Anderson2014}, the covariance matrix computed from the mock catalogues has been corrected using the corresponding Hartlap factors \citep{Hartlap2007} for our number of mocks, number of fitting parameters, and number of bins (see \citealt{Percival2014}). 

\section{BAO fitting results}
\label{sec:fitting}

In this section we present the results of the isotropic and anisotropic BAO fittings in this paper. We compute the correlation functions in 8 $h^{-1}$Mpc bins and consider scales in the range 30 $h^{-1}$Mpc to 180 $h^{-1}$Mpc for the fitting  procedure. This binning size and fitting range is close to optimal \citep[see][in preparation]{Ross2015}.

We fit the correlation functions shown in Figure~\ref{fig:dr12} to a template based on the matter power spectrum for our fiducial cosmology generated by the Boltzmann code \textsc{CAMB} \citep{Lewis2000}. We construct a template in which the correlation function for the fiducial model is convoluted with a Gaussian to reproduce the damping effect on the BAO due to non-linearities \citep{NonLinearBAO2} calibrated on the average of our 1000 mocks (see Section~\ref{sec:mockfitting}). Hereafter this template is designated as the 'de-wiggled' template $\xi^{\textrm{de-wiggled}}$, which we use in our isotropic and anisotropic BAO fittings. We then marginalize over the amplitude and the smooth shape of the correlation function introducing three nuisance polynomial terms and a nuisance amplitude term. The resulting fitting function is therefore: 
\begin{equation}
\xi^{\textrm{fit}}(r)=B_0^2\xi^{\textrm{de-wiggled}}(\alpha r) +A_0 +\frac{A_1}{r} +\frac{A_2}{r^2}
\label{eq:fitmodel}
\end{equation}
where $A_0$, $A_1$ and $A_2$ are parameters that try to capture any smooth deviation from our template due to large-scale systematics and non-linear bias, and $B_0$ is a normalization parameter. All these four coefficients are nuisance parameters which are marginalized over in our fittings. Further details on the fitting methodology used here can be found in \citet[in preparation]{Vargas-Magana2015b}.

In the anisotropic fittings we fit for both $\alpha$ and $\epsilon$ (see e.g. \citealt{Xu2013}), which are related to the angular diameter distance $D_A(z)$ and Hubble parameter $H(z)$.
\begin{equation}
\alpha = \frac{D^{2/3}_A(z)H^{-1/3}(z)/r_{\rm d}}{\left( D^{2/3}_A(z)H^{-1/3}(z)/r_{\rm d} \right)_{\textrm{fid}} } \quad 1+\epsilon = \left( \frac{D_A(z)H(z)}{(D_A(z)H(z))_{\textrm{fid}} } \right)^{-1}
\end{equation}. 

These parameters are related to the dilation factors in the line of sight and the perpendicular directions, $\alpha_{\parallel}$ and $\alpha_{\perp}$, as follows: 
\begin{equation}
\alpha=\alpha_{\parallel}^{1/3}\alpha_{\perp}^{2/3} \qquad 1+\epsilon=\left(\frac{\alpha_{\parallel}}{\alpha_{\perp}}\right)^{1/3}
\end{equation}

which in turn can be written in terms of the angular diameter distance and Hubble parameter via:
\begin{equation}
\alpha_{\perp}=\frac{D_A(z)/r_{\rm d}}{(D_A(z)/r_{\rm d})_{\textrm{fid}}} \qquad \alpha_{\parallel} = \left( \frac{H(z)r_{\rm d}}{(H(z)r_{\rm d})_{\textrm{fid}}}\right)^{-1}
\end{equation}.

The isotropic fittings, which fit for the parameter $\alpha$, can be used to determine the value of the angle-averaged distance $D_V(z)$, which we also report in this paper:
\begin{equation}
D_V(z)=\left[cz(1+z)^2D_A^2(z)H^{-1}(z)\right]^{1/3}
\end{equation}.

A comprehensive analysis of the fitting systematics is presented in \citet[in preparation]{Vargas-Magana2015b}. We refer the reader to that paper for further details.

\begin{figure*}
\centering
\includegraphics[width=0.9\textwidth]{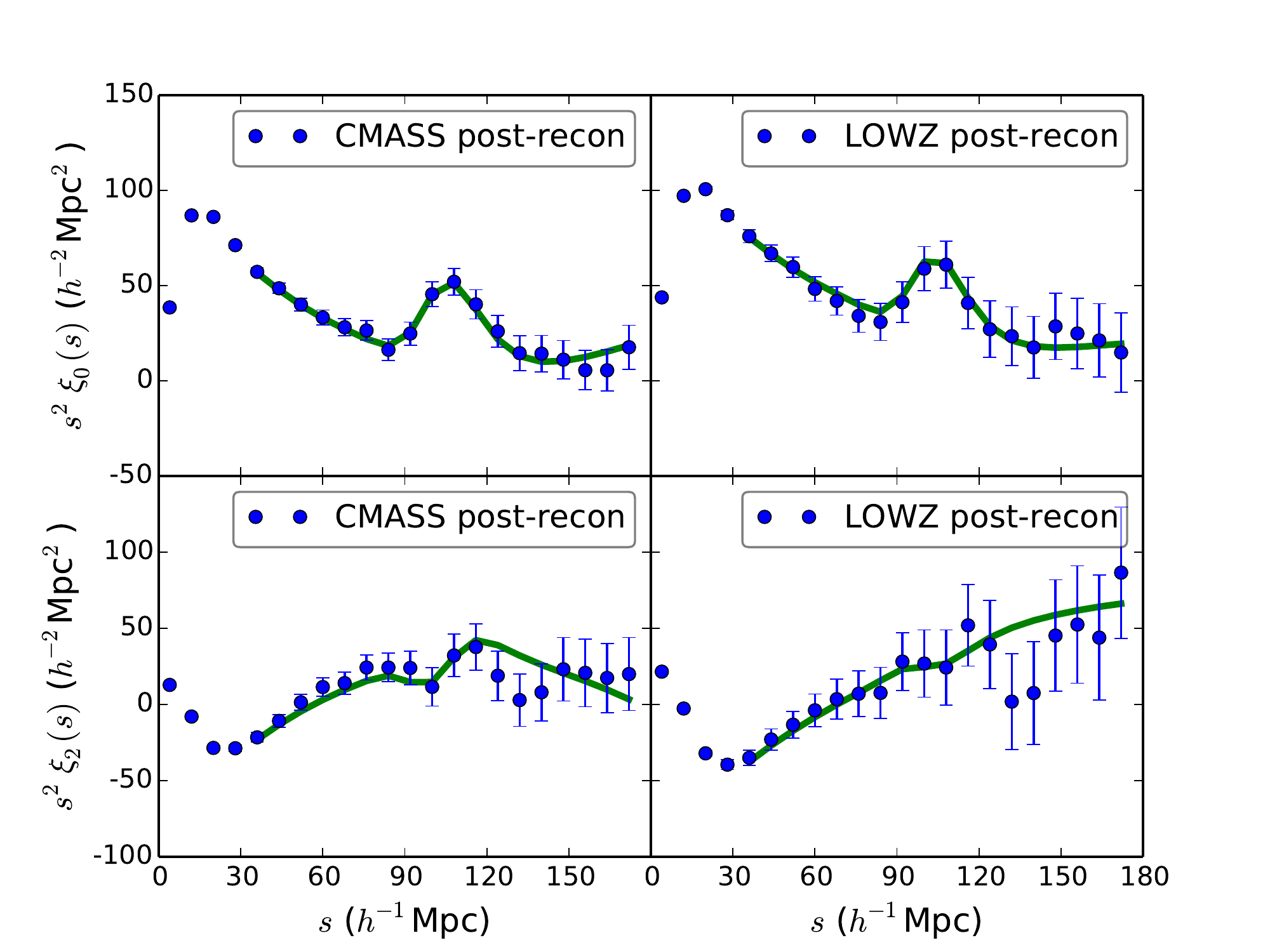}
\caption{The correlation function of CMASS galaxies (left panels) and LOWZ galaxies (right panels). Top panels show the monopole of the correlation function post-reconstruction, bottom panels display the quadrupole of the correlation function. Error bars represent the square root of the diagonal elements of the covariance matrix. In all panels the best-fitting model is presented for reference (solid lines, see text for more details).}
\label{fig:bestfit}
\end{figure*}

Figure~\ref{fig:bestfit} displays the post-reconstruction DR12 correlation functions, for CMASS (left panels) and LOWZ (right panels), together with their best-fitting models. The top panels show the monopole of the correlation function, the bottom panels present the quadrupole. The model of equation~\ref{eq:fitmodel} fits really well the data, with a goodness of fit of $\chi^2=$25 for LOWZ and 26 for CMASS, for 26 degrees of freedom (36 data points minus 10 fitting parameters).

\subsection{Mock fitting results}
\label{sec:mockfitting}
We begin with the results from fitting the QPM mocks. We use 956 mocks for the CMASS sample and 1000 for LOWZ. We compute the median and standard deviations of the geometric parameters $\alpha$, $\epsilon$, $\alpha_{\parallel}$, and $\alpha_{\perp}$ and present them in Table~\ref{tab:fitmocks}. Since our fiducial cosmology corresponds to the input cosmology of the mocks, we expect that on average $\epsilon=0$ and $\alpha=\alpha_{\parallel}=\alpha_{\perp}=1$. Indeed, our recovered parameters values are not biased in any of the pre-reconstruction or post-reconstruction samples. Moreover, the scatter in the measurements provide an approximate idea of the uncertainties we should expect in the data. Post-reconstruction the typical uncertainty in $D_A(z)$ and $H(z)$ is 2.5 per cent and 5.2 per cent respectively for LOWZ, and is 1.6 per cent and 3.1 per cent for CMASS.

\begin{table*}
\centering
\caption{Results of the anisotropic BAO fittings in the QPM mocks of the LOWZ and CMASS samples. We present median values $\tilde{x}$, scatter $S_x$, median uncertainties $\widetilde{\sigma_x}$, and scatter in the uncertainties $S_{\sigma_x}$ for $\alpha_{\parallel}$ and $\alpha_{\perp}$. We also show the median values and scatter for $\alpha$ and $\epsilon$ for reference.  The variables with a tilde indicate the median of that variable. $S$ denotes the root mean square deviation in that variable. \label{tab:fitmocks}}
\begin{tabular}{|l|cccc|cccc|cccc|}
\hline
\hline
                 &  $\widetilde{\alpha}$ & $S_{\alpha}$ & $\widetilde{\epsilon}$ & $S_{\epsilon}$  & $\widetilde{\alpha_{\parallel}}$ & $S_{\alpha_{\parallel}}$ & $\widetilde{\sigma_{\alpha_{\parallel}}}$ & $S_{\sigma_{\alpha_{\parallel}}}$ & $\widetilde{\alpha_{\perp}}$ & $S_{\alpha_{\perp}}$ & $\widetilde{\sigma_{\alpha_{\perp}}}$ & $S_{\sigma_{\alpha_{\perp}}}$ \\
\hline
LOWZ pre-recon   & 1.0036 & 0.0337 & +0.0021 & 0.0362 & 1.0084 & 0.0874 & 0.0905 & 0.0431 & 1.0012 & 0.0406 & 0.0365 & 0.0161 \\
LOWZ post-recon  & 1.0017 & 0.0177 & +0.0009 & 0.0235 & 1.0057 & 0.0526 & 0.0524 & 0.0363 & 1.0007 & 0.0270 & 0.0248 & 0.0089 \\
\hline
CMASS pre-recon  & 1.0025 & 0.0152 & +0.0018 & 0.0196 & 1.0061 & 0.0452 & 0.0482 & 0.0234 & 1.0013 & 0.0217 & 0.0223 & 0.0041 \\
CMASS post-recon & 1.0019 & 0.0105 & +0.0026 & 0.0149 & 1.0067 & 0.0336 & 0.0312 & 0.0168 & 0.9987 & 0.0167 & 0.0156 & 0.0025 \\
\hline
\end{tabular}
\end{table*}

\begin{figure*}
\centering
\includegraphics[width=\textwidth]{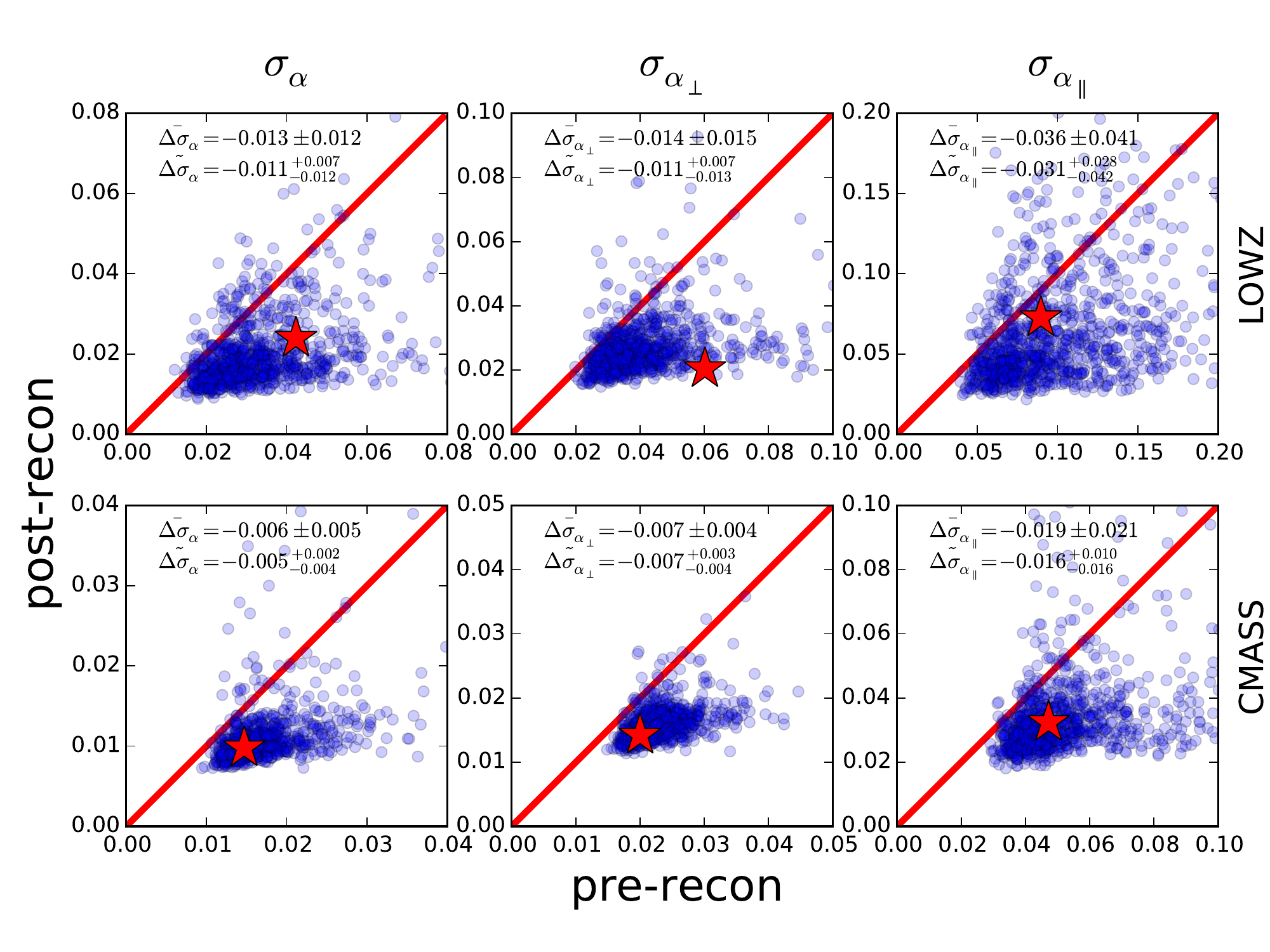}
\caption{Comparison of the pre- and post-reconstruction uncertainties in the anisotropic BAO fittings in QPM mocks. From left to right, we show how reconstruction generally improves $\sigma_{\alpha}$, $\sigma_{\alpha_{\perp}}$ and $\sigma_{\alpha_{\parallel}}$ respectively. The top row presents the results for LOWZ mocks and bottom row shows CMASS mocks. The uncertainties found in the Data Release 12 catalogues are shown with a red star. \label{fig:reconmocks}}
\end{figure*}

\begin{figure}
\centering
\includegraphics[width=0.22\textwidth]{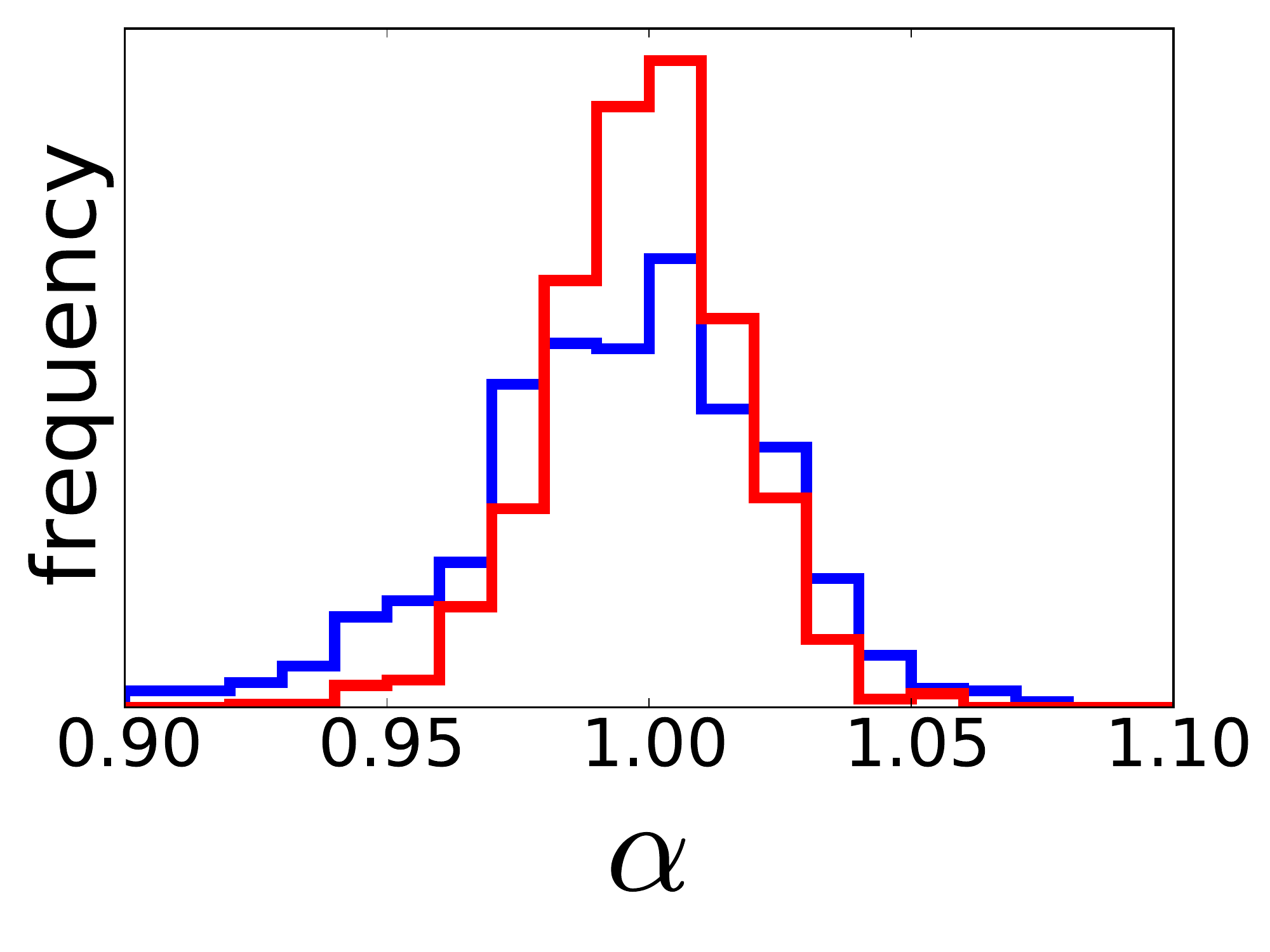}
\includegraphics[width=0.22\textwidth]{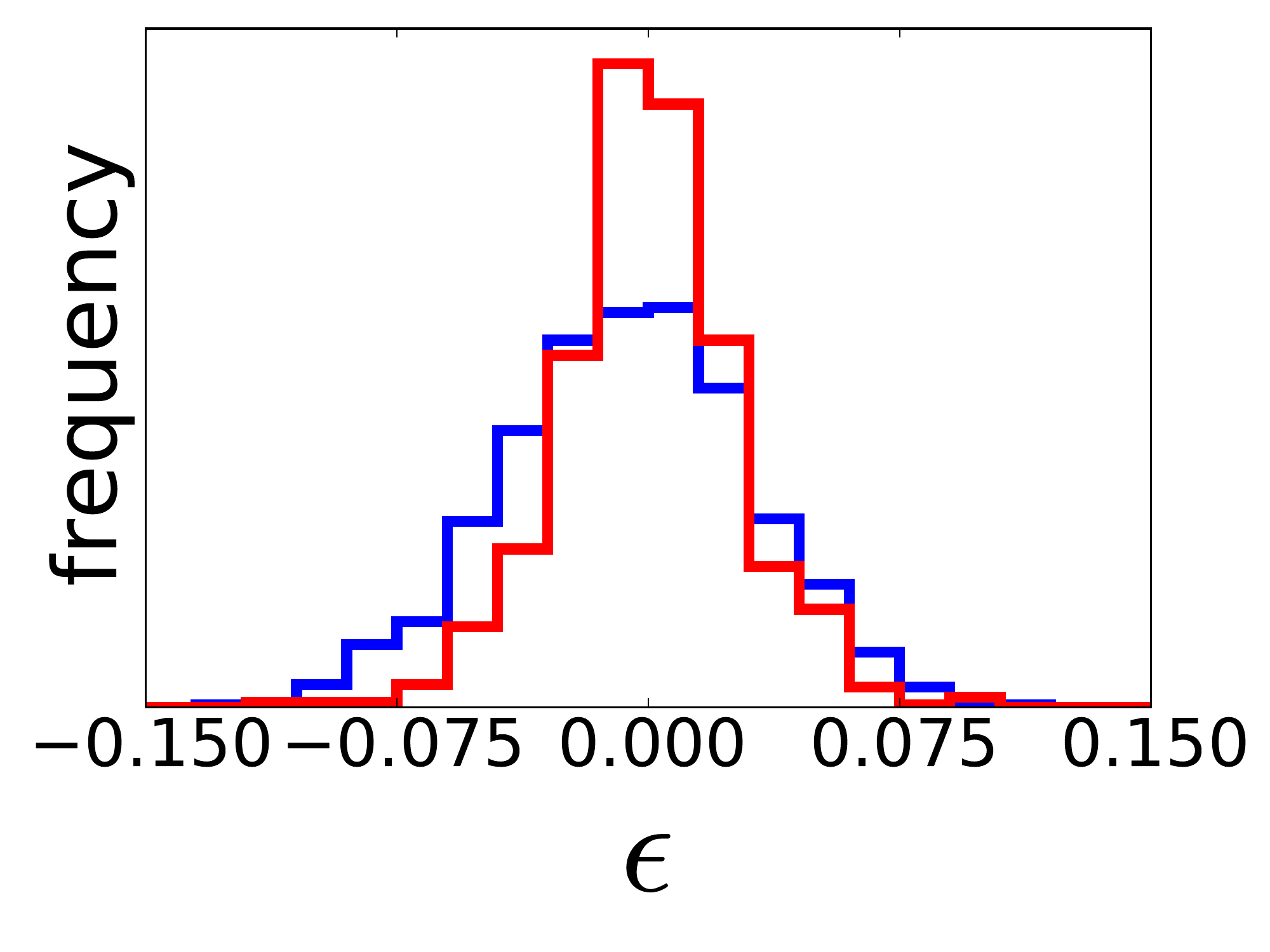}\\
\includegraphics[width=0.22\textwidth]{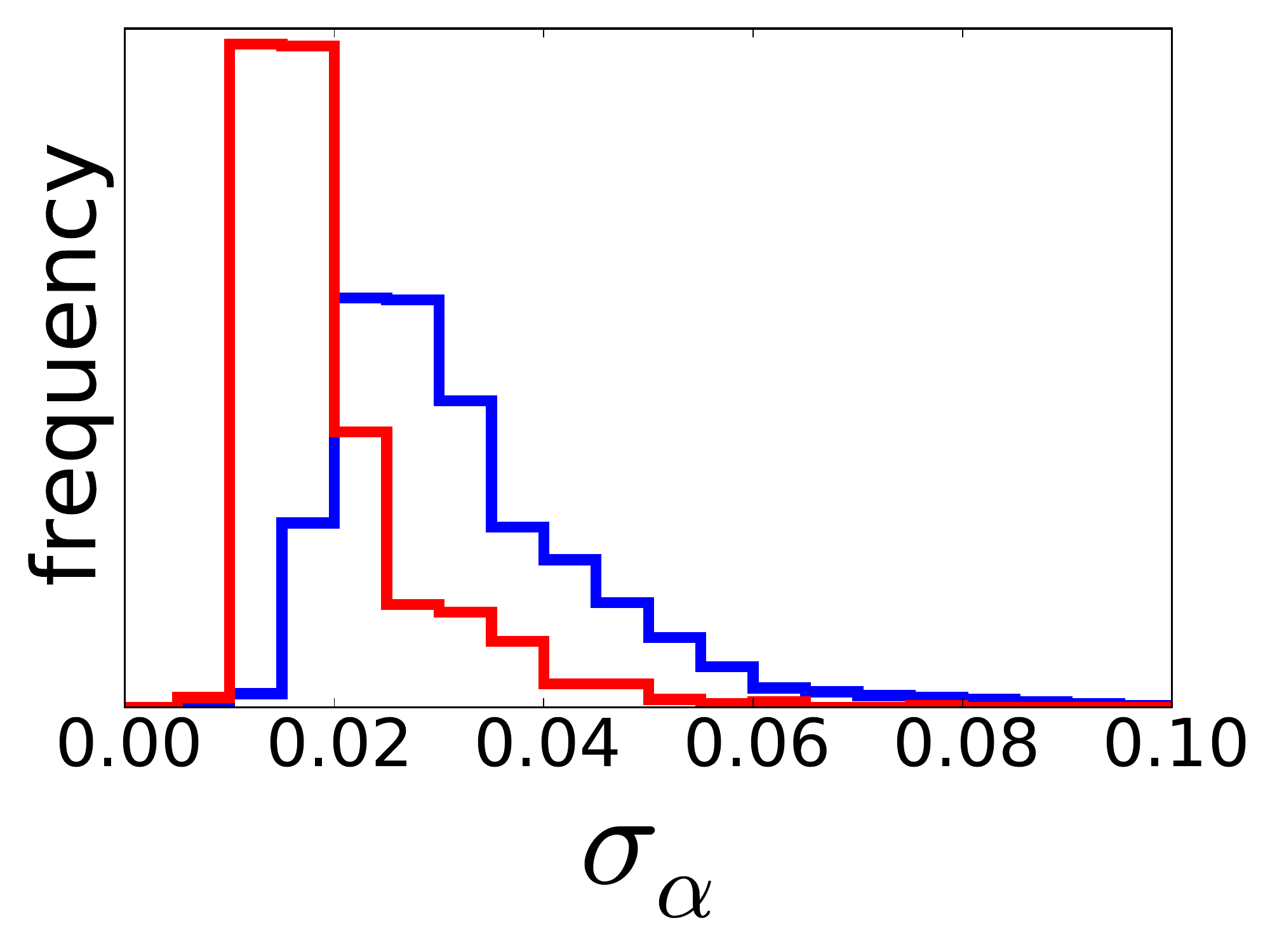}
\includegraphics[width=0.22\textwidth]{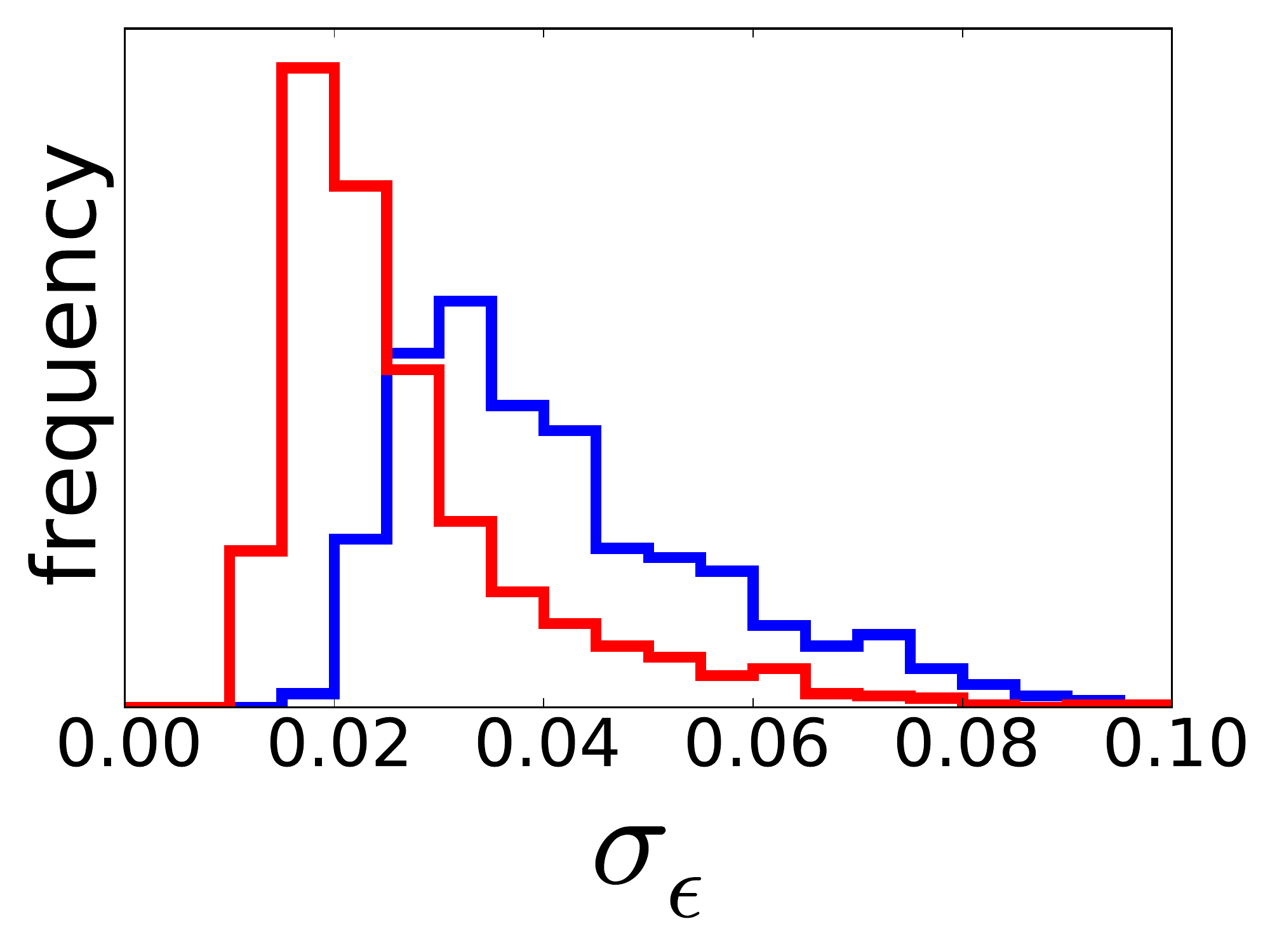}\\
\includegraphics[width=0.22\textwidth]{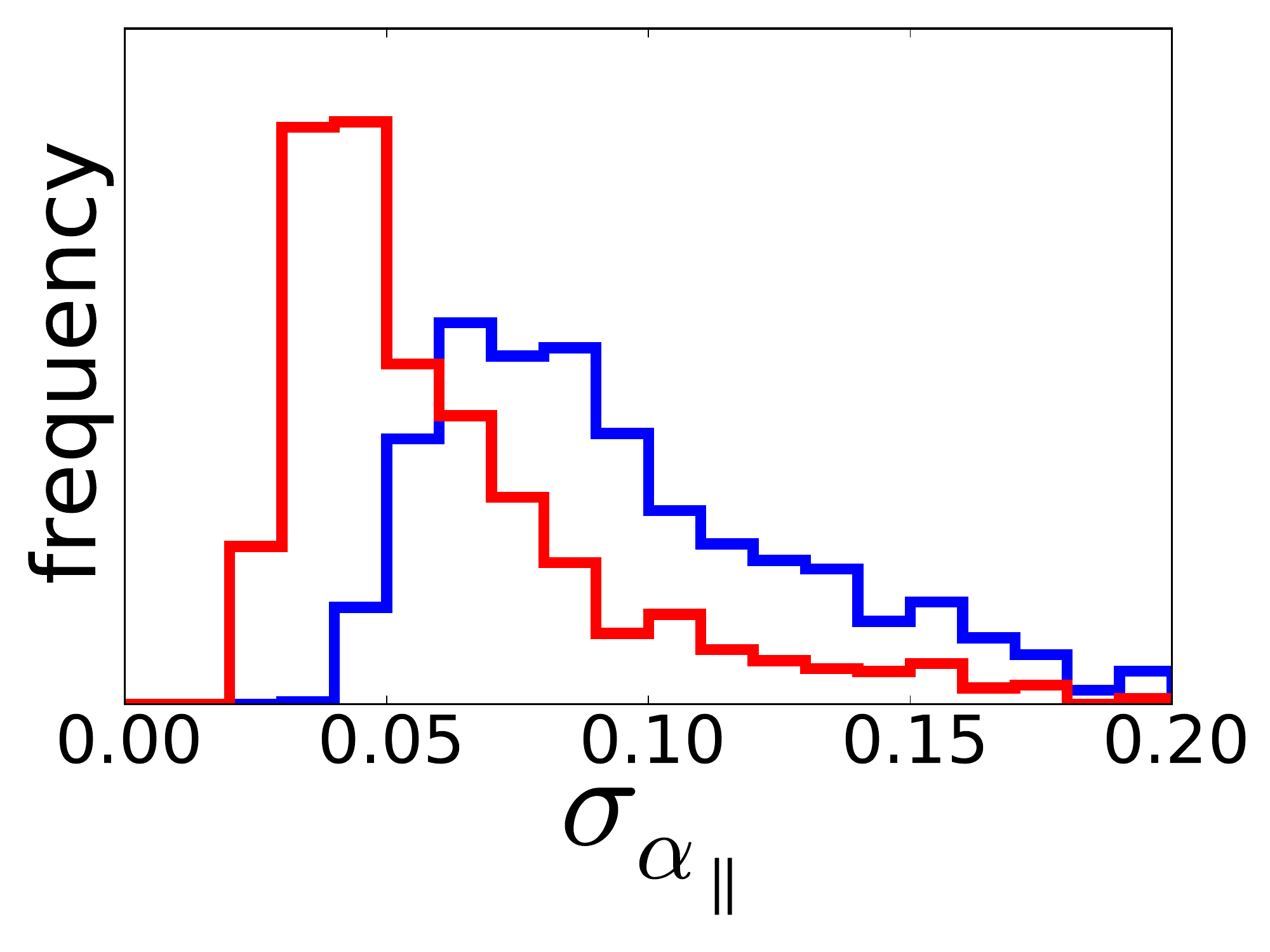}
\includegraphics[width=0.22\textwidth]{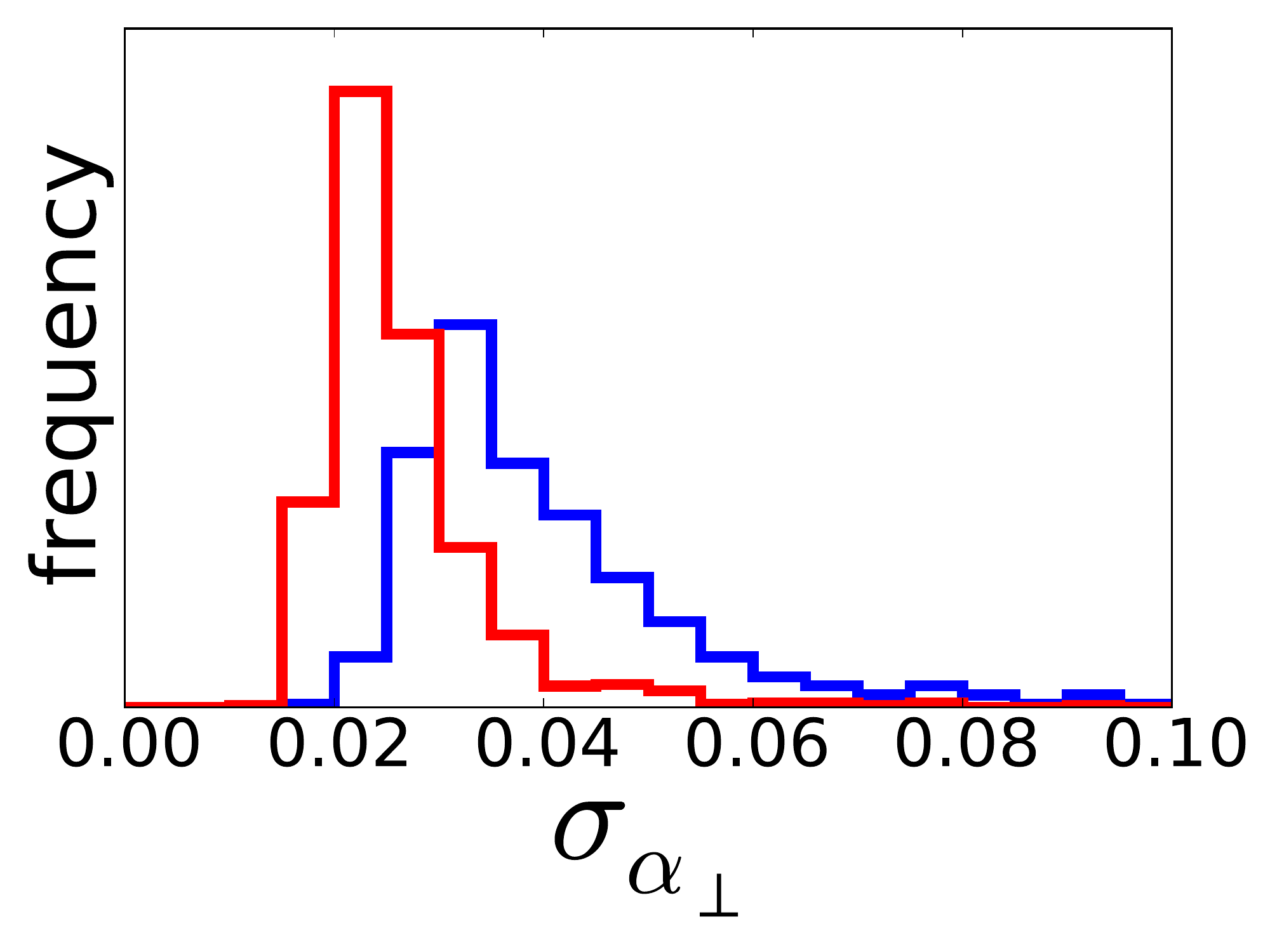}\\
\caption{Statistics of BAO anisotropic fittings in LOWZ mocks. We present the distribution of measured values of $\alpha$ and $\epsilon$ in the mocks (top panels), as well as of their uncertainties $\sigma_{\alpha}$ and $\sigma_{\epsilon}$ (middle panels). Bottom panels show the distribution of the uncertainties in the line of sight distance scale and in the perpendicular direction $\sigma_{\alpha_{\parallel}}$ and $\sigma_{\alpha_{\perp}}$ respectively. Blue lines represent the pre-reconstruction BAO fittings, red lines display the post-reconstruction ones\label{fig:histlowzmocks}.}
\end{figure}

\begin{figure}
\centering
\includegraphics[width=0.22\textwidth]{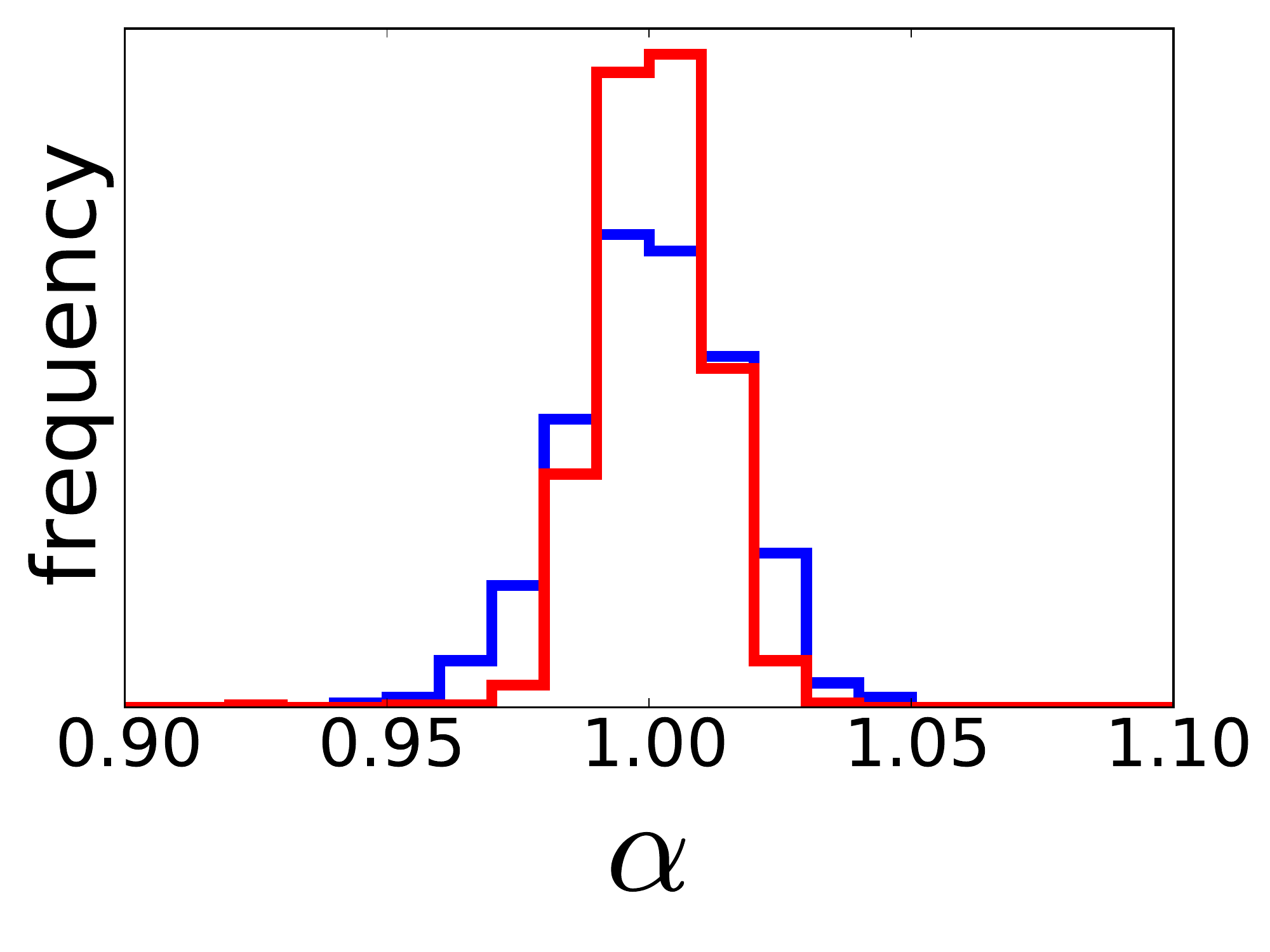}
\includegraphics[width=0.22\textwidth]{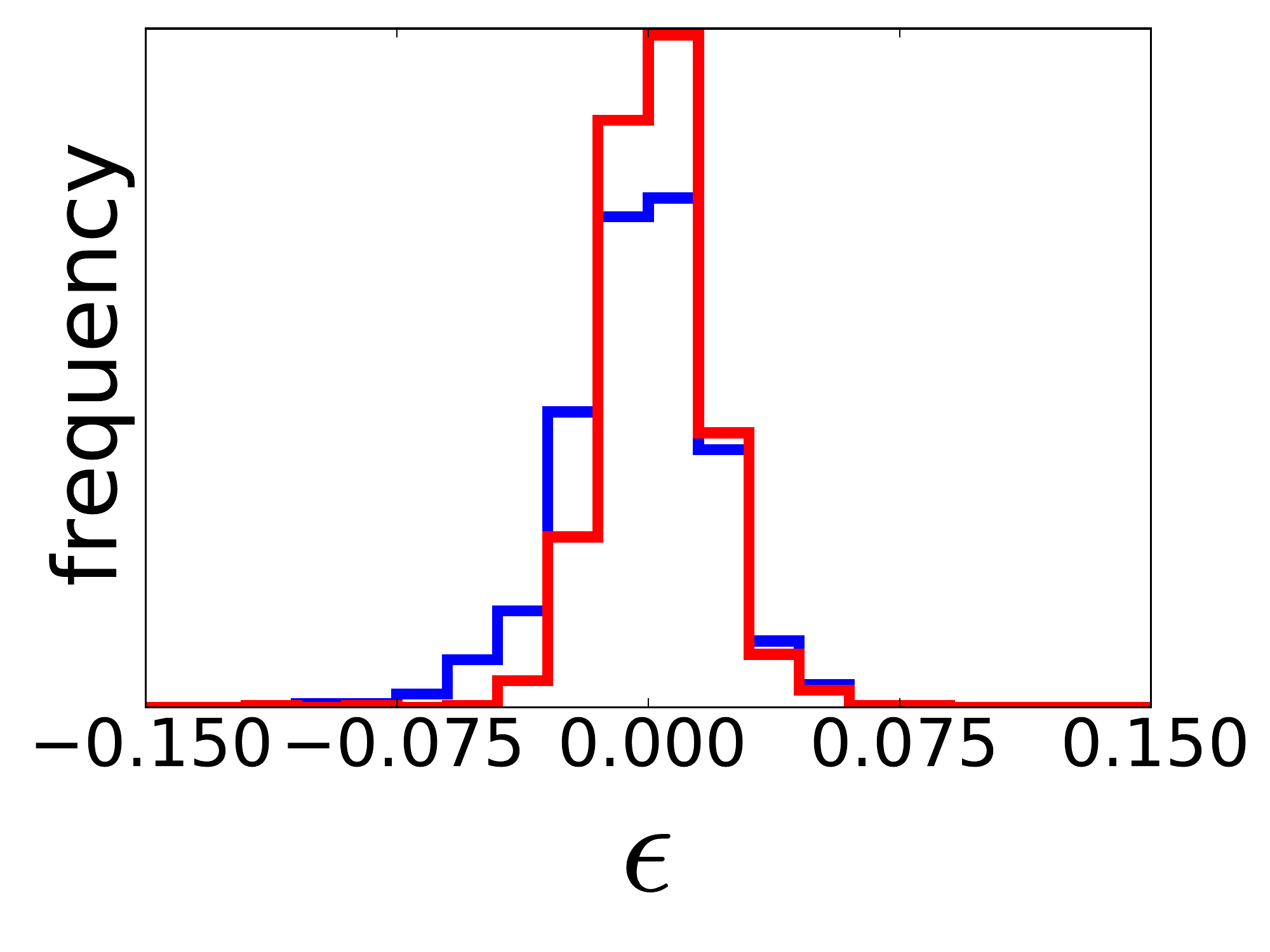}\\
\includegraphics[width=0.22\textwidth]{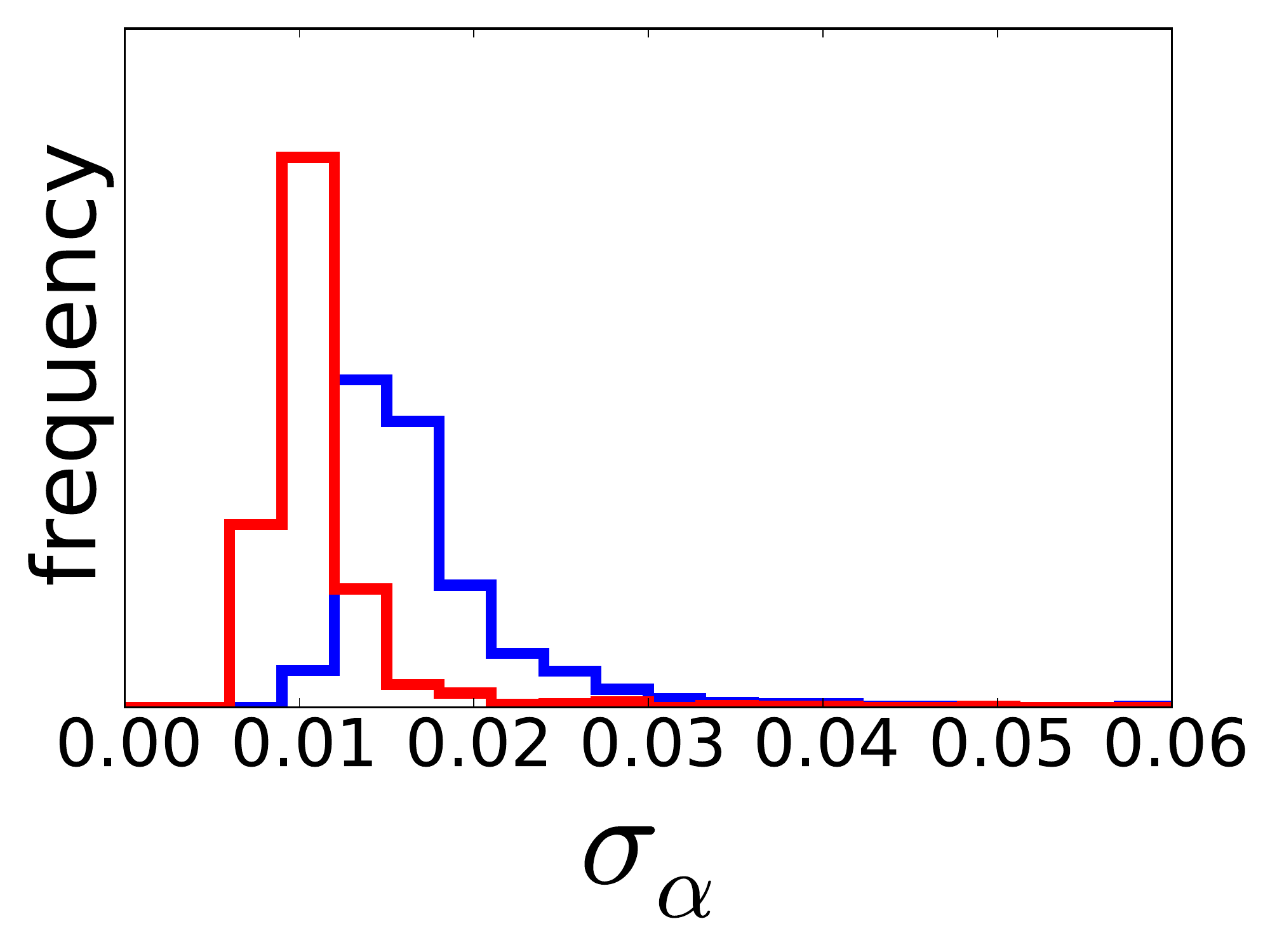}
\includegraphics[width=0.22\textwidth]{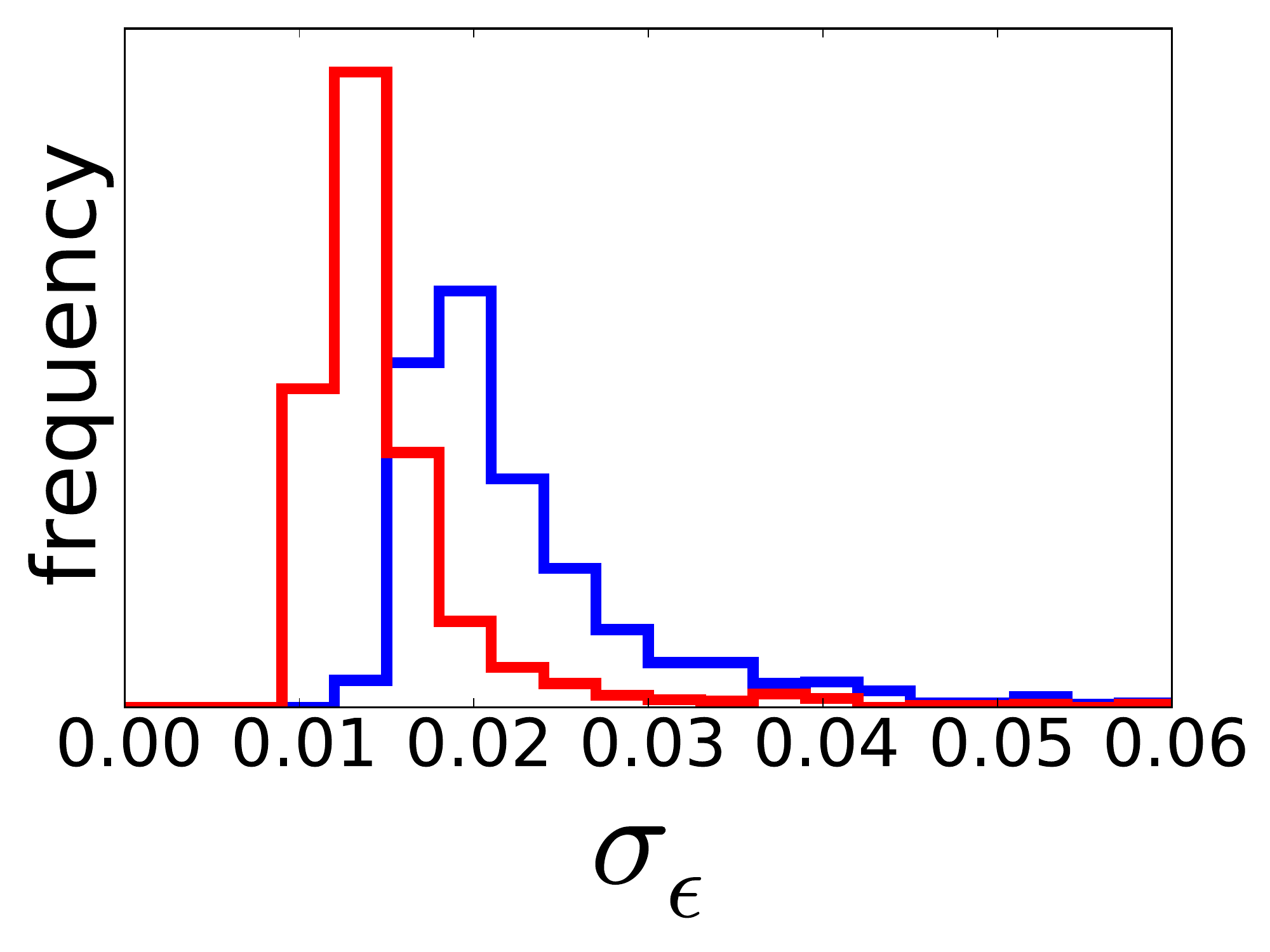}\\
\includegraphics[width=0.22\textwidth]{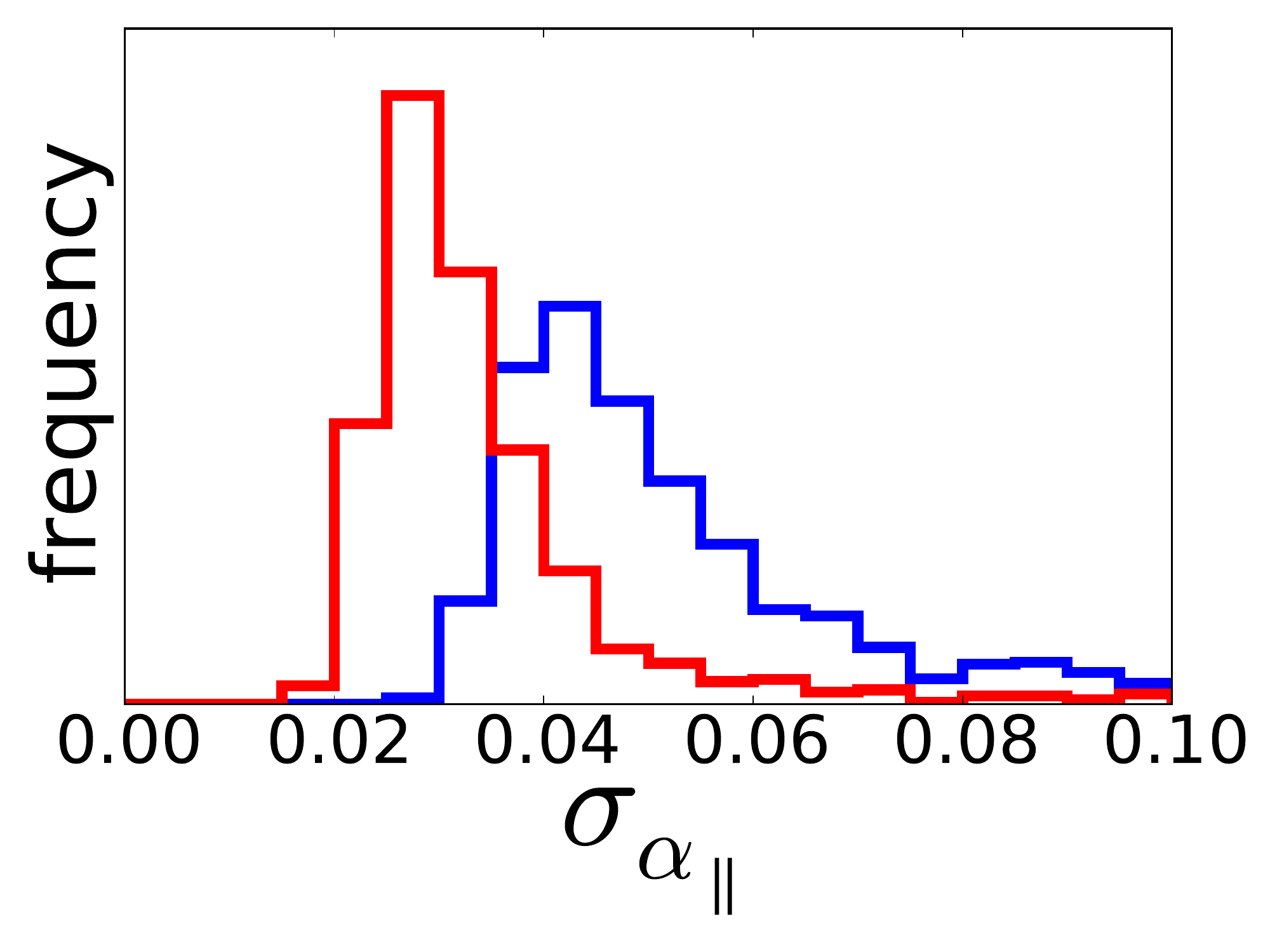}
\includegraphics[width=0.22\textwidth]{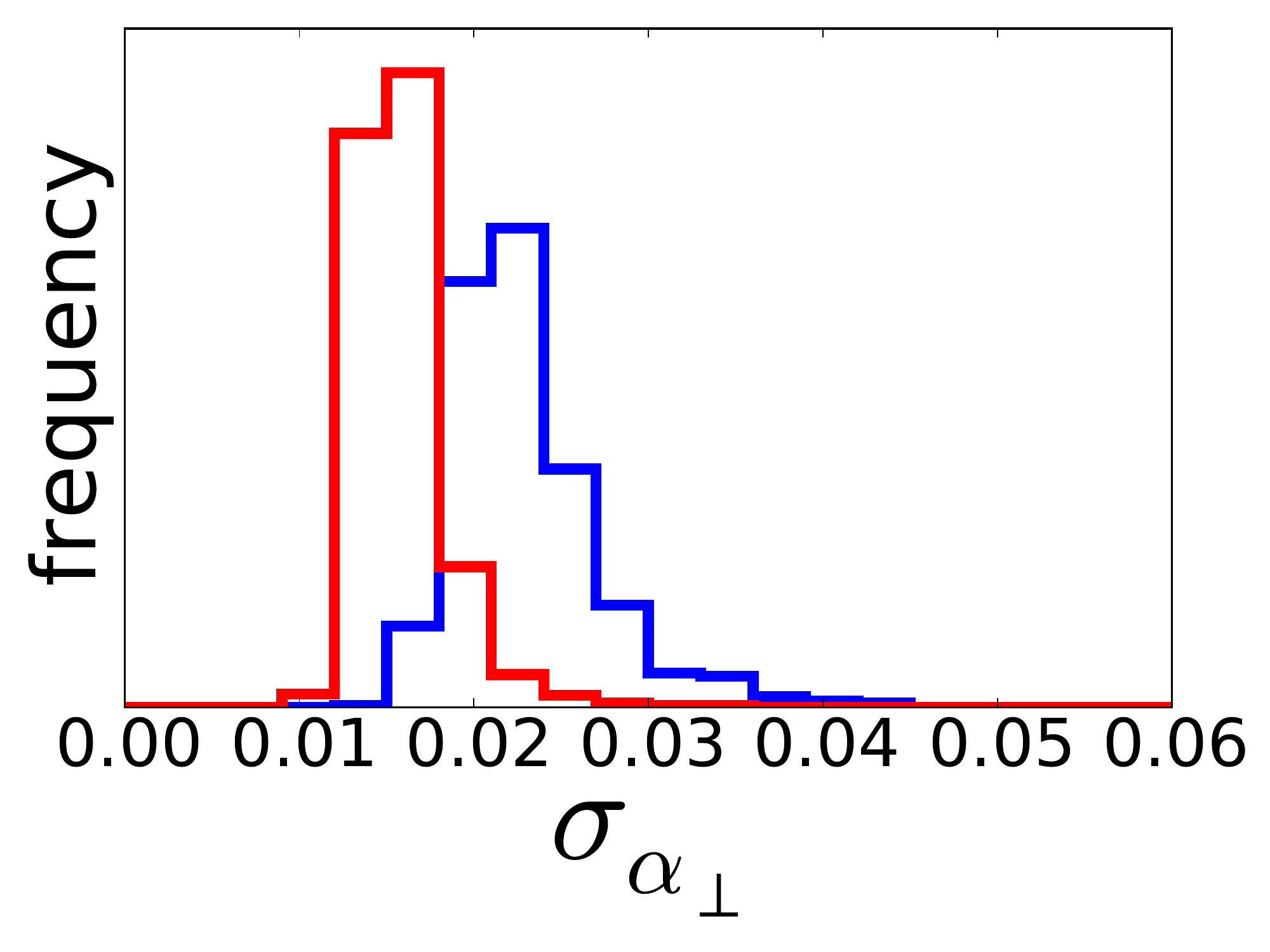}\\
\caption{Statistics of BAO anisotropic fittings in CMASS mocks. We present the distribution of measured values of $\alpha$ and $\epsilon$ in the mocks (top panels), as well as of their uncertainties $\sigma_{\alpha}$ and $\sigma_{\epsilon}$ (middle panels). Bottom panels show the distribution of the uncertainties in the line of sight distance scale and in the perpendicular direction $\sigma_{\alpha_{\parallel}}$ and $\sigma_{\alpha_{\perp}}$ respectively. Blue lines represent the pre-reconstruction BAO fittings, red lines display the post-reconstruction ones\label{fig:histcmassmocks}.}
\end{figure}

The distribution of the uncertainties recovered in the BAO fittings of the mocks are shown in Figure~\ref{fig:histlowzmocks} for LOWZ and Figure~\ref{fig:histcmassmocks} for CMASS, respectively. The blue lines present the distribution before applying density field reconstruction and in red lines the distribution after reconstruction. Overall there is an improvement in the uncertainties of all the geometric parameters for both LOWZ and CMASS samples after reconstruction. A mock-by-mock comparison of the performance of the reconstruction technique on individual mocks can be seen in Figure~\ref{fig:reconmocks}. In this figure each point represents a mock, and its location is given by its uncertainty pre- and post-reconstruction. The points in the region below the red line represent those mocks where reconstruction reduced the uncertainty in the corresponding parameter. In CMASS the vast majority of mocks are improved due to reconstruction, whereas a lower fraction of LOWZ mocks found that improvement. In particular, for $\sigma_{\alpha}$ only 3.2 per cent of the CMASS mocks (8.7 per cent of the LOWZ mocks) were not improved post-reconstruction. For $\sigma_{\alpha_\perp}$ this fraction was 1.8 per cent for CMASS (5.8 per cent for LOWZ), and for $\sigma_{\alpha_\parallel}$ this was 6.9 per cent for CMASS (13.2 per cent for LOWZ).

\subsection{Data fitting results}
We now apply our isotropic and anisotropic fitting analysis to the LOWZ and CMASS Data Release 12 galaxy catalogues. The results are presented in Table~\ref{tab:dr12fit}. Compared to the average values found in the mock catalogues, the constraint on $\alpha_{\perp}$ is better than expected, whereas the constraint on $\alpha_{\parallel}$ is slightly worse. For comparison, we show the Data Release 11 constraints from \cite{Anderson2014} along with our new results in Table~\ref{tab:dr11fit}. There is a slight decrease in the constraining power of the new results for the CMASS sample mainly due to the change in the methodology of generating the mock catalogues. Although the fitting results using MD-Patchy mocks are found to be more constraining than those from QPM mocks, we prefer to err on the conservative side and quote the results from QPM mocks. A comparison of the results with both sets of mocks is revisited in \citet[in preparation]{Anderson2015}.

\begin{table*}
\centering
\caption{Results of the anisotropic fittings of the BAO feature in the correlation function of LOWZ and CMASS galaxy samples, before and after reconstruction. We present the measured value and 1$\sigma$ uncertainties for $\alpha$ and $\epsilon$. Since these two variables are correlated, we include their correlation $\rho_{\alpha\epsilon}=\sigma_{\alpha\epsilon}/\sigma_{\alpha}\sigma_{\epsilon}$. The corresponding values and uncertainties for $\alpha_{\parallel}$ and $\alpha_{\perp}$, together with their correlation, are displayed as well. The last column shows the minimum value of $\chi^2$ and the number of degrees of freedom in the fit.\label{tab:dr12fit}}
\begin{tabular}{|l|ccccccc|}
\hline
\hline
                 &  $\alpha$  & $\epsilon$ & $\rho_{\alpha\epsilon}$ & $\alpha_{\parallel}$ & $\alpha_{\perp}$ & $\rho_{\alpha_{\parallel}\alpha_{\perp}}$ & $\chi^2$/dof \\
\hline
LOWZ pre-recon   & $1.0035 \pm 0.0423$ & $+0.0259 \pm 0.0407$ & -0.1302 & $1.0562 \pm 0.0896$ & $0.9782 \pm 0.0602$ & -0.3034 & 19.2/26 \\
LOWZ post-recon  & $1.0257 \pm 0.0239$ & $+0.0058 \pm 0.0262$ & +0.6770 & $1.0377 \pm 0.0726$ & $1.0197 \pm 0.0204$ & -0.2888 & 24.6/26 \\
\hline
CMASS pre-recon  & $1.0185 \pm 0.0147$ & $-0.0162 \pm 0.0197$ & +0.4127 & $0.9858 \pm 0.0472$ & $1.0352 \pm 0.0200$ & -0.5169 & 26.1/26 \\
CMASS post-recon & $1.0051 \pm 0.0098$ & $-0.0305 \pm 0.0141$ & +0.4169 & $0.9446 \pm 0.0324$ & $1.0368 \pm 0.0142$ & -0.5671 & 25.6/26 \\
\hline
\end{tabular}
\end{table*}

\begin{table*}
\centering
\caption{Comparison of DR11 and DR12 fitting results from the correlation function of LOWZ and CMASS. The value of $\alpha$ from the isotropic fitting is labelled as $\alpha_{\rm{iso}}$ to distinguish it from the value from the anisotropic fitting. The DR11 results are taken from Table 10 of Anderson et al. (2014) and Table 3 of Tojeiro et al. (2014). They correspond to the ones quoted from the correlation function analysis (for a single bin centre choice and de-wiggled template where available), and have been re-scaled to the fiducial cosmology in this paper\label{tab:dr11fit}.}
\begin{tabular}{|l|ccc|ccc|}
\hline
\hline
 & \multicolumn{3}{c}{Pre-Reconstruction} & \multicolumn{3}{c}{Post-Reconstruction} \\
 & $\alpha_{\rm{iso}}$ & $\alpha$ & $\epsilon$ & $\alpha_{\rm{iso}}$ & $\alpha$ & $\epsilon$ \\
 \hline
LOWZ-DR11   &  1.006$\pm$0.033 &        -        &         -        & 1.002$\pm$0.019  &        -        &         -        \\
LOWZ-DR12   &  1.009$\pm$0.030 & 1.004$\pm$0.042 & $+0.026\pm0.041$ & 1.023$\pm$0.017  & 1.026$\pm$0.024 & $+0.006\pm0.026$ \\
\hline
CMASS-DR11   & 1.025$\pm$0.013 & 1.019$\pm$0.014 & $-0.008\pm0.018$ & 1.0145$\pm$0.0090 & 1.0106$\pm$0.0089 & $-0.030\pm0.013$ \\
CMASS-DR12   & 1.015$\pm$0.013 & 1.019$\pm$0.015 & $-0.016\pm0.020$ & 1.0093$\pm$0.0097 & 1.0051$\pm$0.0098 & $-0.031\pm0.014$ \\
\hline
\end{tabular}
\end{table*}

\begin{figure}
\centering
\includegraphics[width=0.45\textwidth]{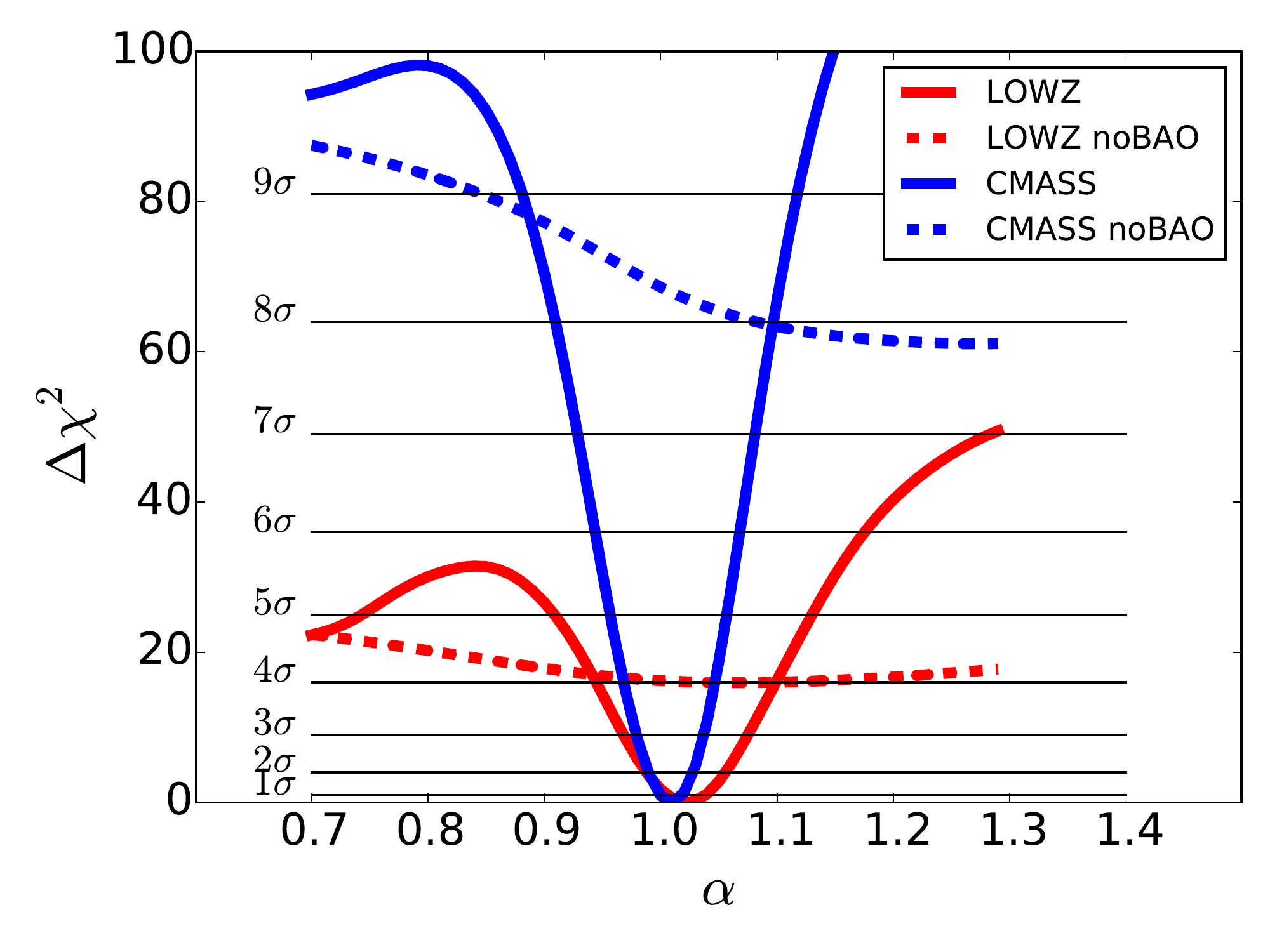}
\caption{Likelihood surfaces $\chi^2(\alpha)$ from the isotropic fitting of the DR12 data post-reconstruction. The results for LOWZ are shown in red and for CMASS are presented in blue. Solid lines correspond to the fitting of the monopole of the correlation function to a template that includes BAO, dashed lines show the case in which the BAO feature in the template has been smoothed away. All lines have been subtracted the $\chi^2$ value at the minimum when the template with BAO is used. For reference, we note the significance of the detection of the BAO with horizontal lines.}
\label{fig:chi2}
\end{figure}

The significance of the BAO detection, however, has increased from Data Release 11 to Data Release 12. Figure~\ref{fig:chi2} presents the $\chi^2$ surface from the isotropic fitting of the DR12 correlation functions. Solid lines represent the difference between $\chi^2(\alpha)$ and its value at the best-fit $\chi^2_{\rm min}$ using our de-wiggled template. Dashed lines show the same when trying to fit the data using a template without a BAO peak \citep{EisensteinHu1998}. The red lines correspond to LOWZ galaxies, blue lines correspond to CMASS galaxies. In this figure the position of the BAO peak is detected with $\simeq 10\sigma$ for CMASS and $\gtrsim 5\sigma$ for LOWZ, whereas the presence of BAO in the correlation function is detected at $\simeq 8\sigma$ for CMASS and $\simeq 4\sigma$ for LOWZ. An apparent decrease in the significance of BAO with increasing $\alpha$ (measured by the difference between the solid and the dashed lines) is seen in this plot. This decrease is caused by our methodology in which the whole de-wiggled template (not just the BAO component) is shifted by $\alpha$ to fit the data, so the ability of our model in equation \ref{eq:fitmodel} to reproduce the shape of the correlation function given the different size of the error bars at different radial separations results in a residual dependence of $\chi^2$ on $\alpha$.

\subsection{Consensus values from $\xi(s)$ and $P(k)$}
This section combines the fitting results from the two-point statistics measuring the clustering in configuration space (i.e. the correlation function, described in this paper) and in Fourier space (i.e. the power spectrum, described in the companion paper \citealt{PK}). As in Data Release 11, these are expected to be highly correlated but sensitive to the clustering in different scale ranges. In configuration space we also take into account that the measurement shows some scatter depending on the position of the bin centres, the resulting value is labelled as \textit{combined} $\xi(s)$. The methodology we follow to combine the results from correlation function and power spectrum was described in Section~4.3 of \cite{Anderson2014}. A brief summary goes as follows: we compute the correlation coefficient $r$ between $\xi(s)$ and $P(k)$ measurements using the fitting results of the mocks, from which we find a correlation of 0.91 for LOWZ and 0.90 for CMASS. The consensus value is then computed as the average of the $\alpha$ measurements from the DR12 $\xi(s)$ and $P(k)$, with an error bar of $\sigma ((1+r)/2)^{1/2}$, where $\sigma$ is the average of the $\sigma_{\alpha}$ values from $\xi(s)$ and $P(k)$. We show in Table~\ref{tab:pkxiqpm} the median and scatter from $\xi(s)$ and $P(k)$ measurements from QPM mocks, as well as the consensus value that combines both. 

\begin{table}
\centering
\caption{Fitting results of the QPM mocks catalogues for LOWZ and CMASS. Power spectrum values taken from the anisotropic fittings from Gil-Mar{\'\i}n et al. (2016)\label{tab:pkxiqpm} and have been re-scaled to the fiducial cosmology in this paper.}
\begin{tabular}{lcccc}
\hline
\hline
Estimator & $\langle \alpha \rangle$ & $S_{\alpha}$ & $\langle \sigma_{\alpha} \rangle$ &$\langle \chi^2 \rangle$ / dof \\
\hline
QPM LOWZ & & & & \\
Consensus $P(k)+\xi(s)$ & 1.00217 & 0.01569 & 0.01617& - \\
combined $\xi(s)$ & 1.00251 & 0.01578 & 0.01624 & - \\
post-recon $P(k)$ & 1.00185 & 0.01631 & 0.01684 & - \\
post-recon $\xi(s)$ & 1.00253 & 0.01617 & 0.01642 & 15.6/15 \\
pre-recon  $\xi(s)$ & 1.00233 & 0.02648 & 0.02750 & 16.1/15 \\
\hline
QPM CMASS & & & & \\
Consensus $P(k)+\xi(s)$ & 0.99981 & 0.00982 & 0.01010 & - \\
combined $\xi(s)$ & 1.00216 & 0.00981 & 0.01025 & - \\
post-recon $P(k)$ & 0.99753 & 0.01034 & 0.01047 & - \\
post-recon $\xi(s)$ & 1.00259 & 0.01016 & 0.01056 & 16.0/15 \\
pre-recon  $\xi(s)$ & 1.00304 & 0.01471 & 0.01528 & 16.1/15 \\
\hline
\end{tabular}
\end{table}

We now apply the same methodology to the observations reported in this paper. The fitting results from Data Release 12 CMASS and LOWZ galaxies are shown in Table~\ref{tab:pkxidr12}. Since the results are highly correlated, the scatter is only slightly reduced due to the combination of both measurements, with respect to the uncertainty on each individual measurement. The good agreement between the BAO fitting analyses in real and Fourier space, being complementary and affected differently by systematic effects, provides a reassuring and robust measurement of the distance scale. In this table we also indicate the consensus results including the systematic error budget we accounted for in \cite{Anderson2014} (labelled as \textit{stat+syst}). This systematic error budget consists of a 0.3 per cent in $\alpha$ for fitting and survey effects, a 0.3 per cent in $\alpha$ for unmodelled astrophysical shifts, and an additional independent systematic error of 0.5 per cent in quadrature to $\epsilon$, as detailed in Section 8.1 of \cite{Anderson2014}.

\begin{table}
\centering
\caption{Results of the fittings of the DR12 data. Power spectrum fittings from Gil-Mar{\'\i}n et al. (2016)\label{tab:pkxidr12} and have been re-scaled to the fiducial cosmology in this paper.}
\begin{tabular}{lcc}
\hline
\hline
Estimator & $\alpha$ & $\chi^2$/dof \\
\hline
DR12 LOWZ & & \\
Consensus $P(k)+\xi(s)$ \textbf{(stat+syst)} & $1.0370\pm0.0177$ & - \\
Consensus $P(k)+\xi(s)$ & $1.0370\pm0.0172$ & - \\
combined $\xi(s)$ & $1.0251\pm0.0169$ & - \\
post-recon $P(k)$ & $1.0489\pm0.0183$ &  56/43\\
post-recon $\xi(s)$ & $1.0230\pm0.0170$ & 9/15 \\
pre-recon $P(k)$   & $1.0061\pm0.0306$ & 36/43 \\
pre-recon $\xi(s)$ & $1.0085\pm0.0300$ & 13/15 \\
\hline
DR12 CMASS & & \\
Consensus $P(k)+\xi(s)$ \textbf{(stat+syst)} & $1.0047\pm0.0099$ & - \\
Consensus $P(k)+\xi(s)$ & $1.0047\pm0.0090$ & - \\
combined $\xi(s)$ & $1.0064\pm0.0096$ & - \\
post-recon $P(k)$ & $1.0029\pm0.0088$ &   37/38 \\
post-recon $\xi(s)$ & $1.0093\pm0.0097$ & 26/15 \\
pre-recon $P(k)$   & $1.0101\pm0.0152$ & 38/38  \\
pre-recon $\xi(s)$ & $1.0153\pm0.0134$ & 12/15 \\
\hline
\end{tabular}
\end{table}

\subsection{Summary of the fitting results}
From the isotropic fittings of the CMASS correlation function, and after accounting for systematic errors, we infer that $D_V(z=0.57)/r_{\rm d}=13.87\pm0.19$ (pre-reconstruction) and $13.79\pm0.14$ (post-reconstruction). Combining with post-reconstruction $P(k)$ BAO fittings, a consensus value of $D_V(z=0.57)/r_{\rm d}=13.73\pm0.14$ is obtained, slightly less constraining than the DR11 measurement of $13.77\pm0.13$. From the isotropic fittings of the LOWZ correlation function we infer that $D_V(z=0.32)/r_{\rm d}=8.47\pm0.25$ (pre-reconstruction) and $8.59\pm0.15$ (post-reconstruction). The combination of the latter with the fitting of BAO in the power spectrum post-reconstruction returns a consensus value of $D_V(z=0.32)/r_{\rm d}=8.71\pm0.15$ which improves the DR11 measurement of $8.47\pm0.17$. The difference between $P(k)$ and $\xi(s)$ measurements is slightly larger in LOWZ than in CMASS, which might explain the difference with the DR11 value despite the small increase in the sampled volume. The post-reconstruction value for the correlation function will be used in our MCMC chains when the isotropic BAO results are included. If the anisotropic BAO results are used instead, then we use the two-dimensional likelihood surface $P(D_A,H)\propto\exp(-0.5\chi^2(D_A,H))$ from the anisotropic fitting without any Gaussian approximation.

We summarize the distance constraints from the anisotropic BAO analysis of the correlation function of the CMASS and LOWZ samples from the Data Release 12 of the Baryon Oscillation Spectroscopic Survey in Table~\ref{tab:distance}. Post-reconstruction, our measurements imply a 1.5 per cent determination of $D_A(z=0.57)$ and a 2.0 per cent measurement of $D_A(z=0.32)$, determining the expansion rate $H(z)$ with a precision of 3.7 per cent at $z=0.57$ and 7.1 per cent at $z=0.32$. The isotropic fittings produce a 1.0 per cent measurement of the distance to redshift $z=0.57$ and a 1.7 per cent determination of the distance to $z=0.32$. All the uncertainties quoted assume that the sound horizon scale is known to within a much better precision and therefore its contribution to the error budget is negligible, which is the case in $\Lambda$CDM and the models studied in Section~\ref{sec:cosmo}, but not necessarily in more general cosmological models. In that case, the uncertainties above would be valid for the quantities $D_A(z)/r_{\rm d}$ and $H(z)r_{\rm d}$, but not for $D_A(z)$ and $H(z)$.

\begin{table*}
\centering
\caption{Distance constraints from the analysis of the BAO in the correlation function of CMASS and LOWZ samples. We quote our results on the angle-averaged distance $D_V(z)$, the angular diameter distance $D_A(z)$, the Hubble parameter $H(z)$, and the correlation $\rho_{D_A,H}$ between $D_A(z)$ and $H(z)$. A fiducial sound horizon value of $r^{\rm fid}_{\rm d}=$147.10 Mpc is assumed. The distance constraints are quoted at redshift $z=0.57$ for the CMASS sample and $z=0.32$ for the LOWZ sample. The error bars in these constraints include the contribution from the systematic error budget. \label{tab:distance}}
\begin{tabular}{lcccc}
\hline
\hline
Sample                     & $D_V(z)r^{\rm fid}_{\rm d}/r_{\rm d}$ & $D_A(z)r^{\rm fid}_{\rm d}/r_{\rm d}$ & $H(z)r_{\rm d}/r^{\rm fid}_{\rm d}$ & $\rho_{D_A,H}$ \\
                           &          (Mpc)                        &       (Mpc)                           &      (km s$^{-1}$Mpc$^{-1}$)        &             \\
\hline
LOWZ Pre-Recon       & $1246\pm37$  &  $941\pm58$ & $77.8\pm6.6$ & 0.31\\
LOWZ Post-Recon      & $1264\pm22$  &  $981\pm20$ & $79.2\pm5.6$ & 0.29\\
\hline
CMASS Pre-Recon      & $2040\pm28$  & $1399\pm28$ & $96.1\pm4.8$ & 0.51\\
CMASS Post-Recon     & $2028\pm21$  & $1401\pm21$ & $100.3\pm3.7$ & 0.55\\
\hline
\end{tabular}
\end{table*}

\begin{figure*}
\includegraphics[width=0.45\textwidth]{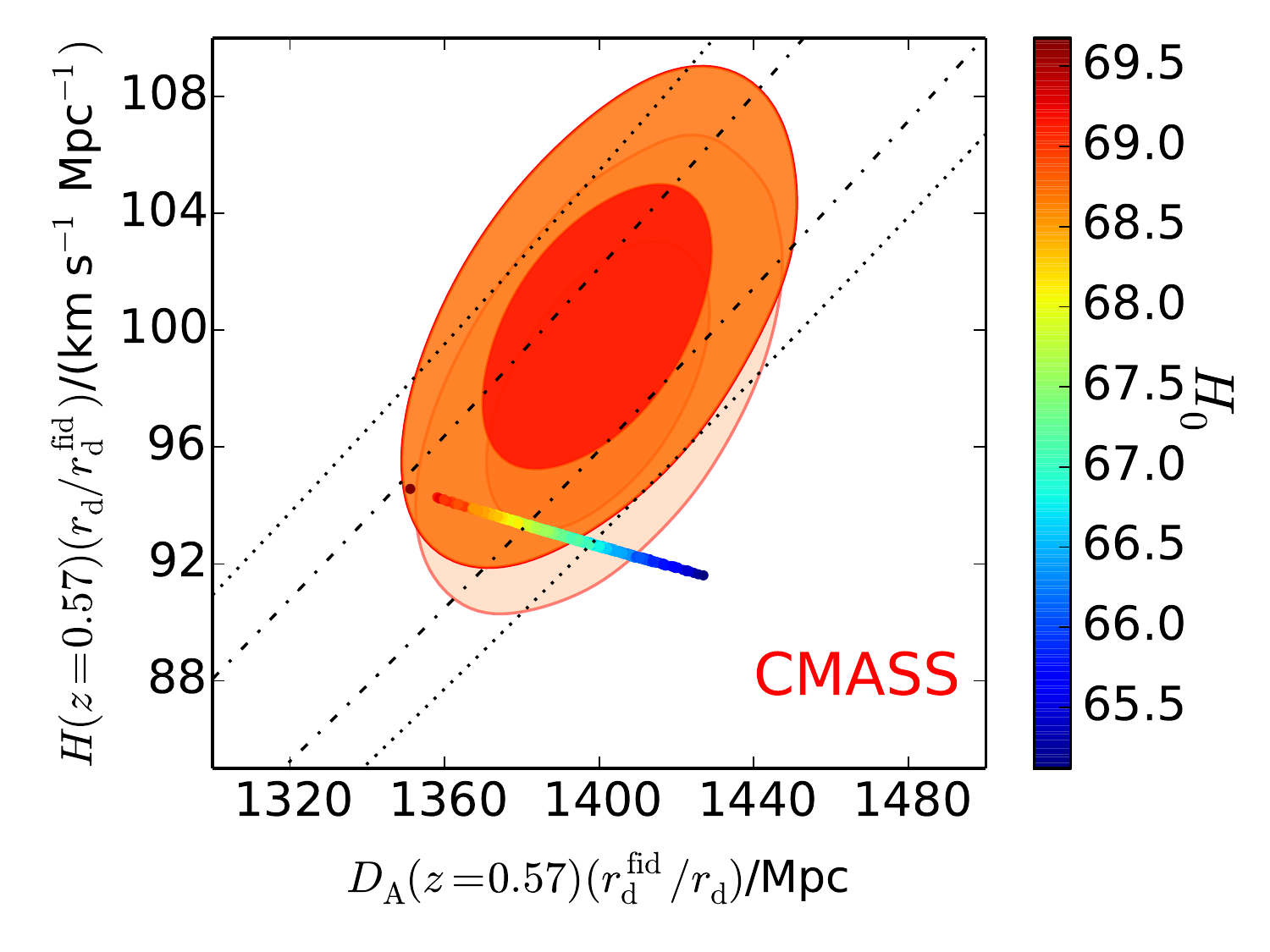}
\includegraphics[width=0.45\textwidth]{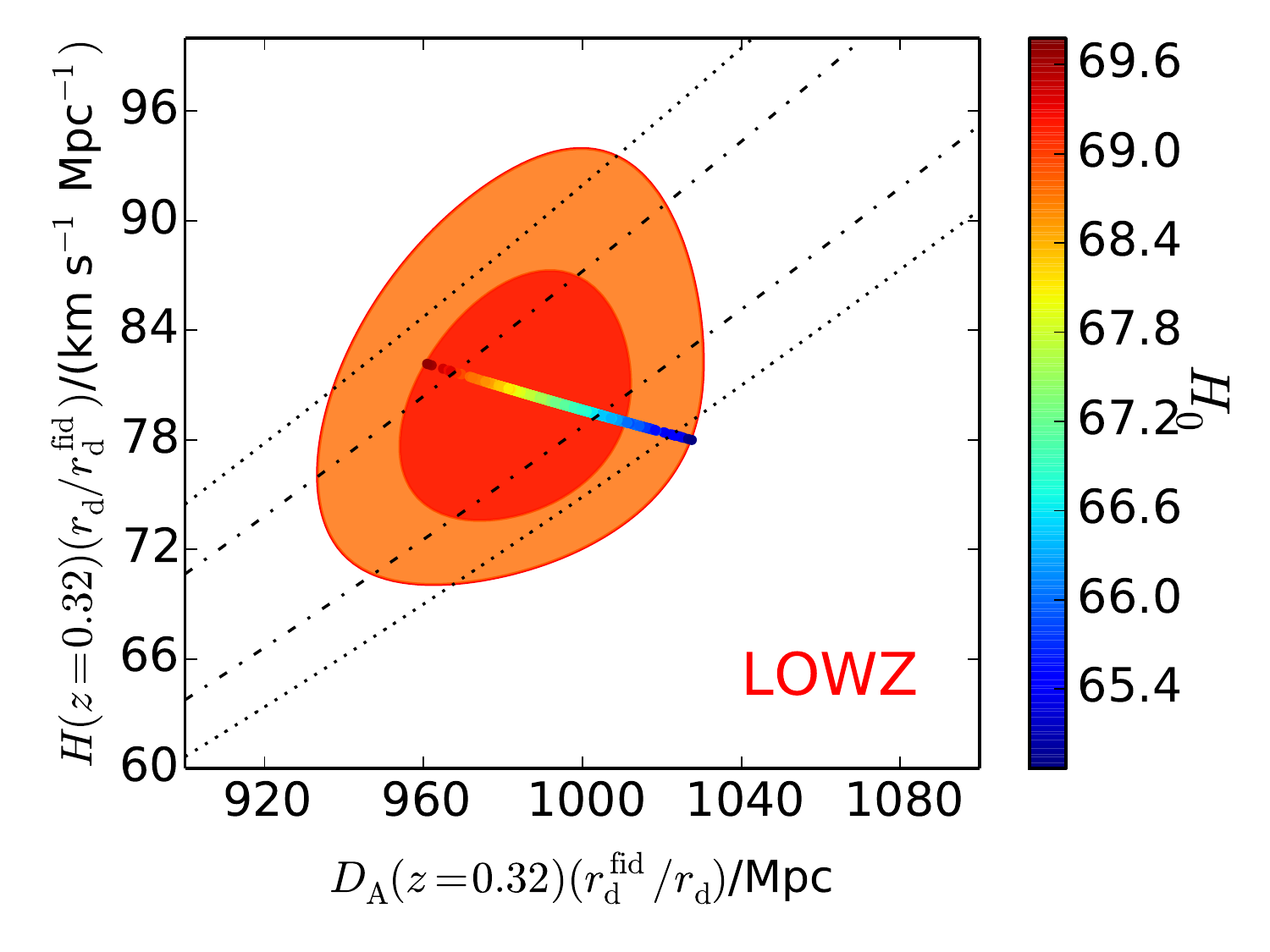}
\caption{Left panel: Anisotropic (solid contours) and isotropic (dashed and dotted lines) constraints for $D_{\rm A}(z=0.57)/r_{\rm d}$ and $H(z=0.57)r_{\rm d}$ from CMASS-DR12 from the analysis using QPM mocks. Light-shaded contours show the constraints from CMASS-DR11 for comparison. Right panel: constraints on $D_{\rm A}(z=0.32)/r_{\rm d}$ and $H(z=0.32)r_{\rm d}$ from the isotropic and anisotropic fitting of LOWZ-DR12 also using QPM mocks. In both cases 1$\sigma$ and 2$\sigma$ contours are shown. Also included is a small region colour-coded according to the value of the Hubble constant from the constraints from Planck 2015 temperature and polarization power spectrum data assuming a $\Lambda$CDM model.}
\label{fig:dah}
\end{figure*}

Figure~\ref{fig:dah} compares the distance constraints from the post-reconstruction isotropic and the anisotropic analysis. Contours represent the anisotropic fitting of CMASS and LOWZ and dashed lines the isotropic fitting. In both cases we show the 1-sigma and 2-sigma constraints. The faint contours in the left panel show the DR11 CMASS constraints for comparison. We also include, for reference, the constraints from a Planck 2015 $\Lambda$CDM model \citep{Planck2015Cosmo} colour-coded according to the corresponding value of the Hubble constant (colour points). We note that the DR12 CMASS contours have shifted slightly upwards compared to DR11, favouring larger values of $H(z=0.57)r_{\rm d}/r^{\rm fid}_{\rm d}$. However, this change is not large enough to make our measurement inconsistent with the $\Lambda$CDM model prediction from Planck 2015.

\section{Cosmological interpretation}
\label{sec:cosmo}
In this section we infer the constraints from the BOSS Data Release 12 BAO measurements on the cosmological parameters from different cosmological models. We use the Markov Chain Monte Carlo (MCMC) code \textsc{CosmoMC}\footnote{http://cosmologist.info/cosmomc} \citep{Lewis2002} to compute our cosmological constraints. Our goal in this section is double. First, we present a comparison with the results in \cite{Anderson2014} with the same cosmological datasets except for the DR12 BAO measurements reported in this paper, addressing the improvement from our updated BAO measurements on constraining the cosmological parameters. Second, we want to take advantage of more powerful cosmological datasets not available at the time of \cite{Anderson2014} to obtain updated cosmological constraints that can be compared with current literature, such as \cite{Planck2015Cosmo}.

\begin{figure*}
\includegraphics[width=0.4\textwidth]{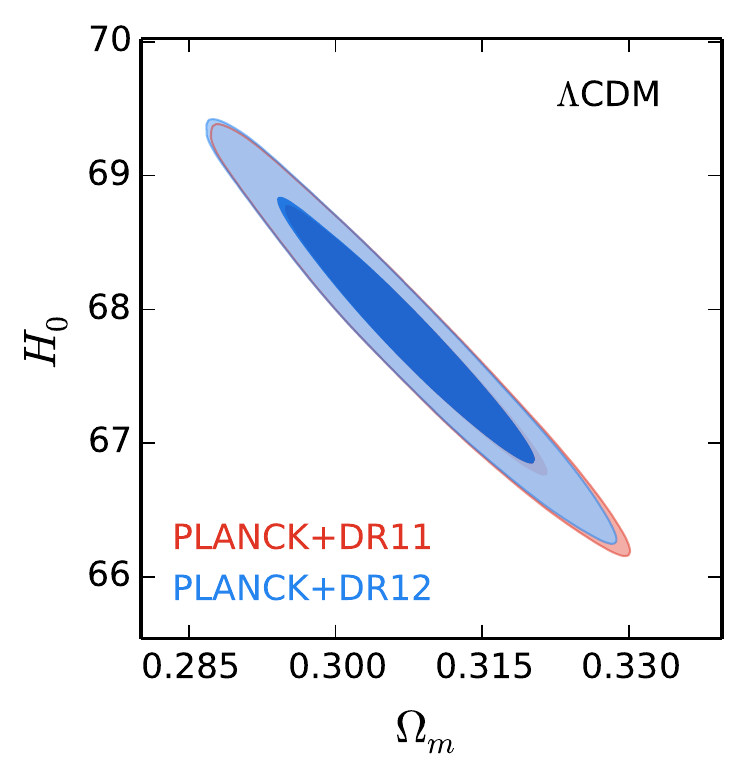}
\includegraphics[width=0.4\textwidth]{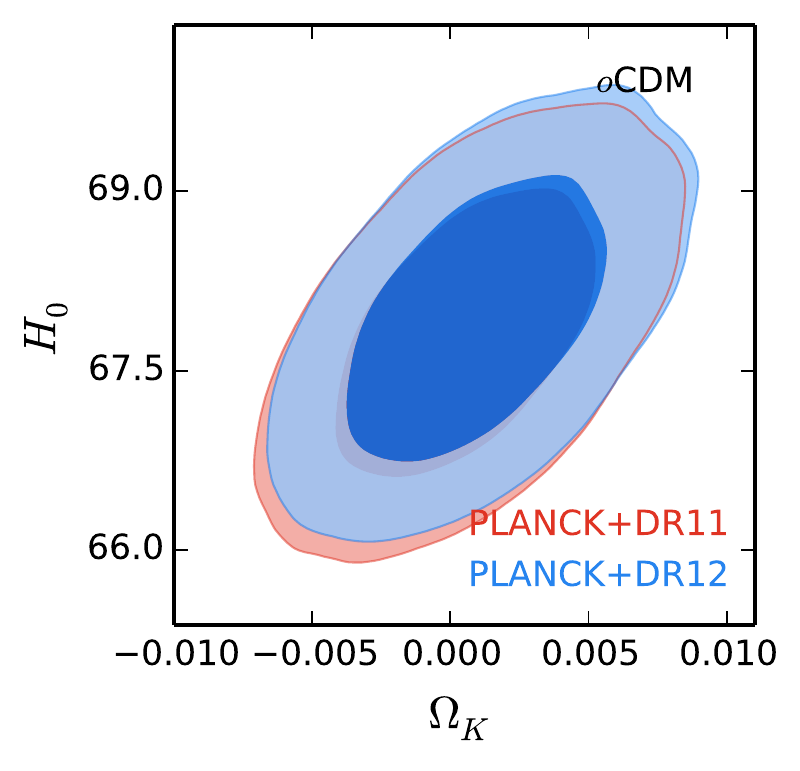}
\\
\includegraphics[width=0.4\textwidth]{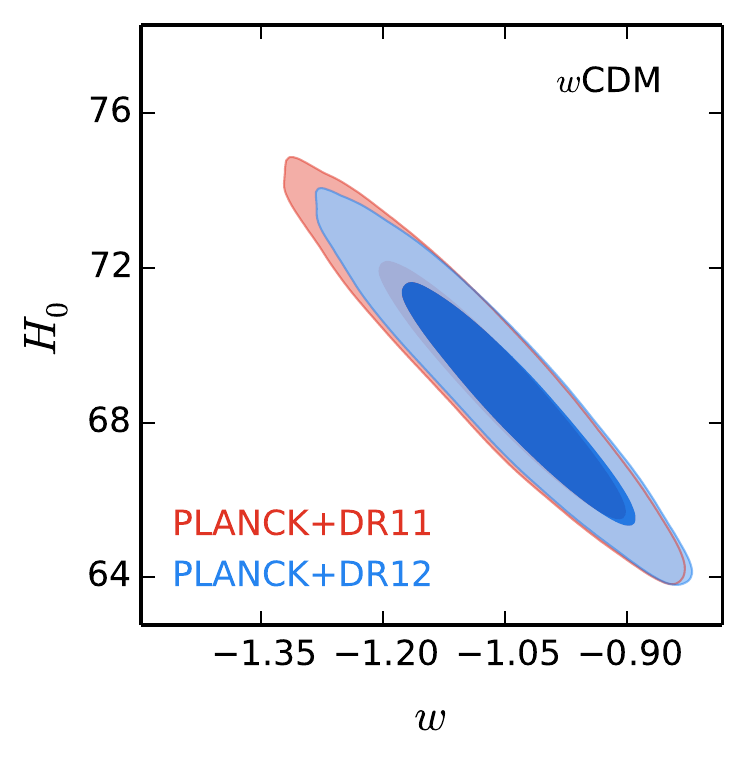}
\includegraphics[width=0.4\textwidth]{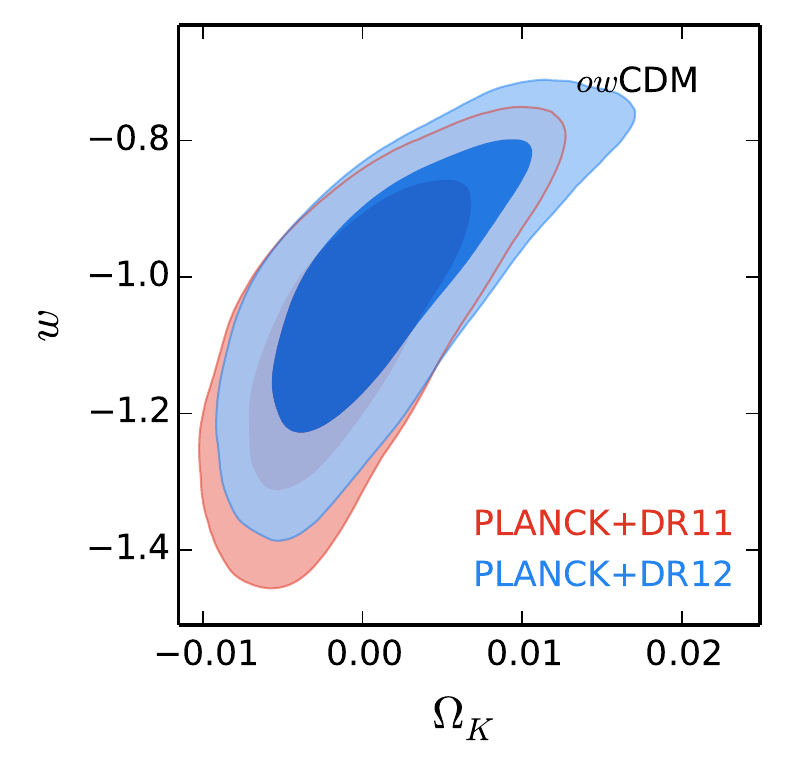}
\\
\includegraphics[width=0.4\textwidth]{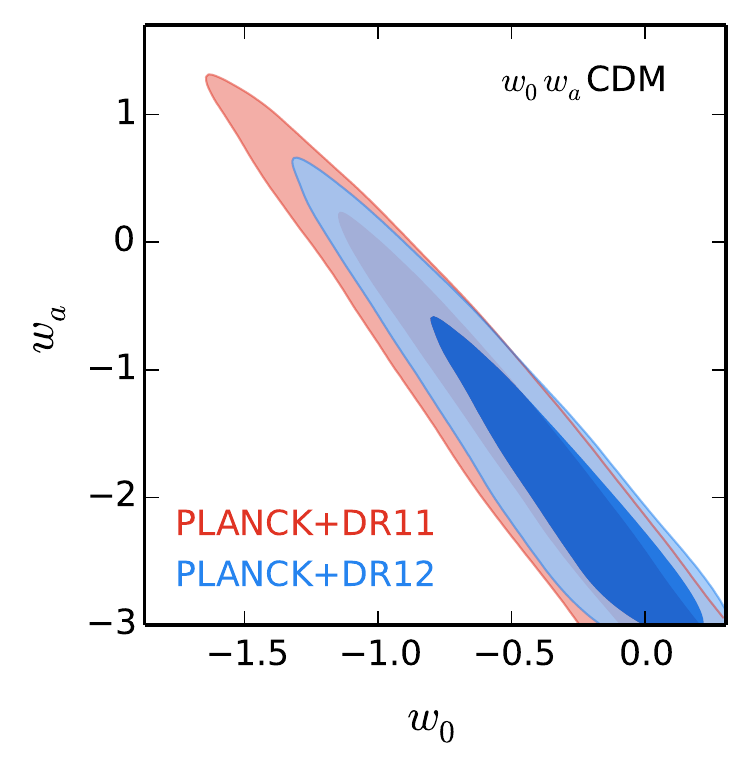}
\includegraphics[width=0.4\textwidth]{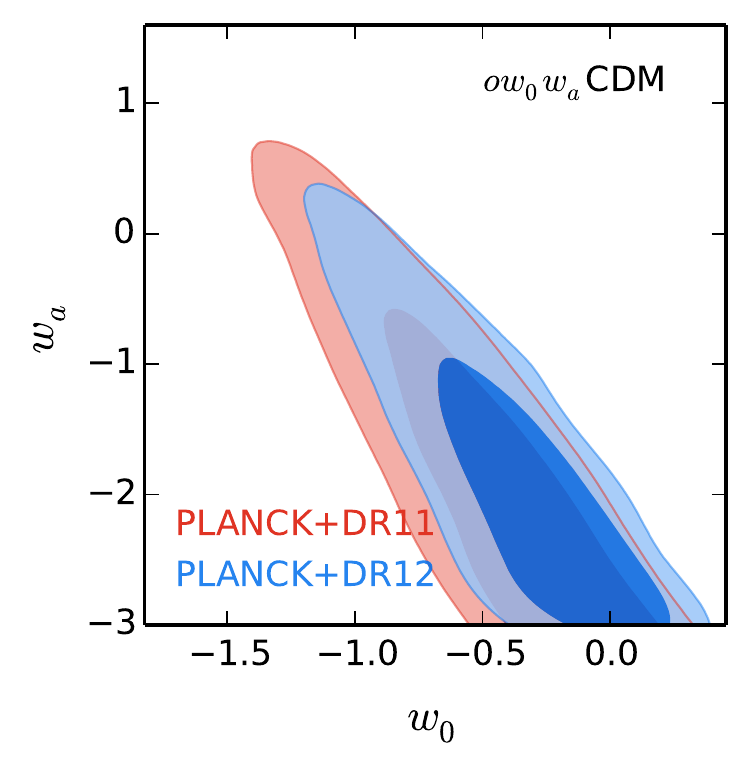}
\caption{Cosmological constraints for different cosmological models, using a combination of Planck13 CMB data and our DR12 BAO measurements (blue contours). Also shown for comparison are the constraints from Anderson et al. (2014) where the DR11 BAO measurements are used instead (red contours). Left panels assume a flat Universe: $\Lambda$CDM (top panel), $w$CDM (middle panel), and $w_0w_a$CDM (bottom panel). On the right panels the curvature is a free parameter: $o$CDM (top panel), $ow$CDM (middle panel), and $ow_0w_a$CDM (bottom panel). The following priors are assumed: $-0.1<\Omega_k<+0.1$, $-3<w_0<+1$, and $-3<w_a<+3$.}
\label{fig:cosmology}
\end{figure*}

We begin with a comparison to the results in \cite{Anderson2014}. Here we combine our BAO measurements with Cosmic Microwave Background (CMB) data from Planck+WP \citep{Planck2013}, which hereafter we refer to as Planck13, or simply as \textit{Planck}. The current CMB measurements from the Planck satellite \citep{Planck2015Overview} are referred to as Planck15. Figure~\ref{fig:cosmology} displays the cosmological constraints from Planck13 and our Data Release 12 BAO constraints (blue contours) compared to those from the same CMB data combined with Data Release 11 BAO (red contours). The results are shown for the different cosmological models studied here: flat Universe (left panels) where dark energy is described by a cosmological constant ($\Lambda$CDM, top panel), dark energy with constant but arbitrary equation of state ($w$CDM, middle panel), and with a time-dependent equation of state ($w_0w_a$CDM, bottom panel). Also shown are their non-flat versions where curvature is a free parameter (right panels: o$\Lambda$CDM, o$w$CDM, o$w_0w_a$CDM respectively). The following priors are assumed: $-0.1<\Omega_k<+0.1$, $-3<w_0<+1$, and $-3<w_a<+3$. As readily seen in this plot, the change from DR11 is incremental. Furthermore, we have checked that the small decrease in the size of these contours is mostly driven by the smaller error bar in the DR12 LOWZ distance constraint.  

A full compilation of our cosmological results is presented in Table~\ref{tab:cosmo}. In these MCMC chains, when we include the Type-1a Supernovae data we use the Union 2 compilation by the Supernovae Cosmology Project \citep{Union} for direct comparison with \cite{Anderson2014}. We include the low-redshift BAO measurement from the 6-degree field galaxy redshift survey (6DF) \citep{Beutler2011}, and the Lyman-$\alpha$ BAO measurements from \cite{Delubac2014} and \cite{Font2014} \footnote{For Data Release 11, an extensive study of the cosmological consequences of Galaxy and Lyman-$\alpha$ BAO from BOSS was presented in \cite{JOINTBAO}.}.

\begin{table*}
\centering
\caption{Cosmological constraints by different dataset combinations in the cosmological models $\Lambda$CDM, oCDM, $w$CDM, o$w$CDM, $w_0w_a$CDM, and o$w_0w_a$CDM. For direct comparison to Table 15 in Anderson et al. (2014), we compare the cosmological constraints from combining Planck13 with distance scale measurements from BOSS DR12 galaxies as well as lower and higher redshift BAO measurements from the 6-degree field galaxy redshift survey (6DF) and the BOSS-Lyman alpha forest (Ly$\alpha$F), respectively. We also compare how these combinations benefit from the constraining power of type-Ia Supernovae from the Union 2 compilation by the Supernovae Cosmology Project (SN). The WMAP and \textit{e}WMAP cases have been added for comparison. 'CMASS-iso' indicates the isotropic measurement from the CMASS sample, whereas the anisotropic one is referred to simply as 'CMASS'. 'LOWZ' is the isotropic measurement from the LOWZ sample. 'BAO' stands for the combination CMASS + LOWZ + 6DF + Ly$\alpha$F. Numbers in parenthesis represent the uncertainty in the accompanying value, e.g. 0.123 (45) should be read as $0.123\pm0.045$.\label{tab:cosmo}}
\begin{tabular}{llllllll}
\hline
\hline
Cosmological & Data Sets & $\Omega_{\rm m} h^{2}$ & $\Omega_{\rm m}$ & $H_{0}$ & $\Omega_{\rm K}$ & $w_{0}$ & $w_{a}$ \\
Model & & & & km s$^{-1}$ Mpc$^{-1}$ & & & \\
\hline
$\Lambda$CDM & Planck + CMASS-iso + LOWZ & 0.1413 (14) & 0.307 (8) & 67.9 (6) & \nodata & \nodata & \nodata \\
$\Lambda$CDM & Planck + CMASS + LOWZ & 0.1413 (13) & 0.307 (8) & 67.8 (6) & \nodata & \nodata & \nodata \\
$\Lambda$CDM & Planck + BAO & 0.1416 (13) & 0.309 (8) & 67.7 (6) & \nodata & \nodata & \nodata \\
$\Lambda$CDM & Planck + CMASS + LOWZ + SN & 0.1412 (13) & 0.307 (8) & 67.9 (6) & \nodata & \nodata & \nodata \\
$\Lambda$CDM & Planck + BAO + SN & 0.1415 (13) & 0.308 (7) & 67.7 (5) & \nodata & \nodata & \nodata \\
$\Lambda$CDM & WMAP + BAO + SN & 0.1398 (22) & 0.301 (8) & 68.2 (7) & \nodata & \nodata & \nodata \\
$\Lambda$CDM & \textit{e}WMAP + BAO + SN & 0.1409 (16) & 0.300 (8) & 68.5 (6) & \nodata & \nodata & \nodata \\
\hline
oCDM & Planck + CMASS-iso + LOWZ & 0.1418 (25) & 0.307 (8) & 68.0 (8) & +0.0008 (30) & \nodata & \nodata \\
oCDM & Planck + CMASS + LOWZ & 0.1421 (25) & 0.308 (8) & 67.9 (7) & +0.0010 (30) & \nodata & \nodata \\
oCDM & Planck + BAO & 0.1424 (25) & 0.310 (8) & 67.8 (7) & +0.0010 (29) & \nodata & \nodata \\
oCDM & Planck + CMASS + LOWZ + SN & 0.1418 (24) & 0.307 (8) & 68.0 (7) & +0.0008 (29) & \nodata & \nodata \\
oCDM & Planck + BAO + SN & 0.1420 (24) & 0.308 (7) & 67.9 (7) & +0.0008 (29) & \nodata & \nodata \\
oCDM & WMAP + BAO + SN & 0.1387 (41) & 0.299 (9) & 68.1 (7) & -0.0015 (40) & \nodata & \nodata \\
oCDM & \textit{e}WMAP + BAO + SN & 0.1367 (34) & 0.296 (8) & 68.0 (7) & -0.0050 (35) & \nodata & \nodata \\
\hline
$w$CDM & Planck + CMASS-iso + LOWZ & 0.1429 (22) & 0.290 (19) & 70.3 (26) & \nodata & -1.11 (11) & \nodata \\
$w$CDM & Planck + CMASS + LOWZ & 0.1420 (21) & 0.301 (15) & 68.7 (20) & \nodata & -1.04 (9) & \nodata \\
$w$CDM & Planck + BAO & 0.1415 (21) & 0.309 (13) & 67.7 (17) & \nodata & -1.00 (7) & \nodata \\
$w$CDM & Planck + CMASS + LOWZ + SN & 0.1422 (19) & 0.300 (12) & 68.9 (15) & \nodata & -1.05 (7) & \nodata \\
$w$CDM & Planck + BAO + SN & 0.1419 (19) & 0.305 (11) & 68.2 (14) & \nodata & -1.02 (6) & \nodata \\
$w$CDM & WMAP + BAO + SN & 0.1371 (35) & 0.308 (11) & 66.7 (16) & \nodata & -0.92 (8) & \nodata \\
$w$CDM & \textit{e}WMAP + BAO + SN & 0.1372 (28) & 0.313 (11) & 66.3 (15) & \nodata & -0.88 (7) & \nodata \\
\hline
o$w$CDM & Planck + CMASS-iso + LOWZ & 0.1419 (24) & 0.282 (28) & 71.3 (36) & -0.0019 (40) & -1.17 (18) & \nodata \\
o$w$CDM & Planck + CMASS + LOWZ & 0.1422 (25) & 0.307 (22) & 68.2 (24) & +0.0016 (49) & -1.01 (13) & \nodata \\
o$w$CDM & Planck + BAO & 0.1423 (25) & 0.320 (18) & 66.8 (18) & +0.0034 (46) & -0.94 (10) & \nodata \\
o$w$CDM & Planck + CMASS + LOWZ + SN & 0.1421 (25) & 0.301 (14) & 68.8 (16) & +0.0001 (35) & -1.05 (8) & \nodata \\
o$w$CDM & Planck + BAO + SN & 0.1423 (25) & 0.308 (13) & 68.0 (14) & +0.0010 (34) & -1.01 (7) & \nodata \\
o$w$CDM & WMAP + BAO + SN & 0.1372 (43) & 0.308 (13) & 66.7 (16) & +0.0000 (46) & -0.92 (8) & \nodata \\
o$w$CDM & \textit{e}WMAP + BAO + SN & 0.1356 (35) & 0.308 (13) & 66.4 (14) & -0.0028 (42) & -0.91 (7) & \nodata \\
\hline
$w_0w_a$CDM & Planck + CMASS-iso + LOWZ & 0.1431 (22) & 0.333 (48) & 66.2 (52) & \nodata & -0.68 (46) & -1.13 (114) \\
$w_0w_a$CDM & Planck + CMASS + LOWZ & 0.1425 (21) & 0.370 (37) & 62.3 (34) & \nodata & -0.34 (34) & -1.83 (86) \\
$w_0w_a$CDM & Planck + BAO & 0.1423 (20) & 0.373 (29) & 61.9 (26) & \nodata & -0.31 (28) & -1.90 (75) \\
$w_0w_a$CDM & Planck + CMASS + LOWZ + SN & 0.1430 (22) & 0.307 (17) & 68.3 (19) & \nodata & -0.94 (19) & -0.42 (63) \\
$w_0w_a$CDM & Planck + BAO + SN & 0.1428 (22) & 0.314 (16) & 67.5 (17) & \nodata & -0.89 (18) & -0.48 (61) \\
$w_0w_a$CDM & WMAP + BAO + SN & 0.1367 (42) & 0.304 (16) & 67.1 (17) & \nodata & -0.97 (16) & 0.12 (56) \\
$w_0w_a$CDM & \textit{e}WMAP + BAO + SN & 0.1363 (31) & 0.303 (15) & 67.2 (17) & \nodata & -1.00 (15) & 0.33 (41) \\
\hline
o$w_0w_a$CDM & Planck + CMASS-iso + LOWZ & 0.1419 (25) & 0.326 (46) & 66.5 (50) & -0.0043 (45) & -0.65 (41) & -1.61 (104) \\
o$w_0w_a$CDM & Planck + CMASS + LOWZ & 0.1417 (24) & 0.368 (37) & 62.3 (33) & -0.0017 (52) & -0.33 (31) & -1.97 (80) \\
o$w_0w_a$CDM & Planck + BAO & 0.1420 (24) & 0.374 (29) & 61.7 (25) & -0.0003 (49) & -0.29 (26) & -1.94 (75) \\
o$w_0w_a$CDM & Planck + CMASS + LOWZ + SN & 0.1420 (25) & 0.309 (17) & 67.9 (18) & -0.0030 (45) & -0.86 (20) & -0.85 (86) \\
o$w_0w_a$CDM & Planck + BAO + SN & 0.1422 (25) & 0.315 (16) & 67.3 (17) & -0.0013 (43) & -0.86 (19) & -0.66 (78) \\
o$w_0w_a$CDM & WMAP + BAO + SN & 0.1368 (44) & 0.304 (16) & 67.2 (18) & +0.0033 (71) & -0.99 (17) & 0.27 (68) \\
o$w_0w_a$CDM & \textit{e}WMAP + BAO + SN & 0.1357 (35) & 0.304 (15) & 66.9 (17) & -0.0012 (56) & -0.97 (16) & 0.18 (56) \\
\hline
\end{tabular}
\end{table*}

This table reveals that the combination of CMB+BAO+SN greatly improves the constraints on curvature and dark energy. There is an improvement in the Figure of Merit of dark energy \citep{DETF} of 10 per cent with respect to our CMB+BAO+SN results from Data release 11 and a 50 per cent improvement if the SN sample is replaced by the recent JLA compilation from \cite{Betoule2014}.

Now we study the constraints from the BAO in CMASS and LOWZ combined with the recent Planck15 temperature plus polarization power spectrum \citep{Planck2015Likelihood}; these results are shown in Table~\ref{tab:planck2015}. In order to present the most up-to-date results we also replace the Union 2 Supernovae sample with the more recent JLA compilation \citep{Betoule2014}. The low-redshift BAO measurement of 6DF here is combined with that from the SDSS Main Galaxy Sample (MGS) sample \citep{MGS}.

\begin{table*}
\centering
\caption{Cosmological constraints from Planck15+LOWZ+CMASS and from Planck15+LOWZ+CMASS+MGS+6DF+JLA. 'CMASS' indicates the anisotropic measurement from the CMASS sample, whereas 'LOWZ' is the isotropic measurement from the LOWZ sample. 'BAO' stands for the combination LOWZ + CMASS + MGS + 6DF. Numbers in parenthesis represent the uncertainty in the accompanying value, e.g. 0.123 (45) should be read as $0.123\pm0.045$.\label{tab:planck2015}}
\begin{tabular}{llllllll}
\hline
Cosmological & Data Sets & $\Omega_{\rm m} h^{2}$ & $\Omega_{\rm m}$ & $H_{0}$ & $\Omega_{\rm K}$ & $w_{0}$ & $w_{a}$ \\
Model & & & & km s$^{-1}$ Mpc$^{-1}$ & & & \\
\hline
$\Lambda$CDM & Planck15 + LOWZ + CMASS & 0.1418 (9) & 0.310 (6) & 67.7 (4) & \nodata & \nodata & \nodata \\
$\Lambda$CDM & Planck15 + BAO + SN & 0.1420 (9) & 0.311 (6) & 67.6 (4) & \nodata & \nodata & \nodata \\
\hline
oCDM & Planck15 + LOWZ + CMASS & 0.1424 (13) & 0.308 (6) & 68.0 (6) & +0.0012 (19) & \nodata & \nodata \\
oCDM & Planck15 + BAO + SN & 0.1424 (13) & 0.310 (6) & 67.8 (6) & +0.0008 (20) & \nodata & \nodata \\
\hline
$w$CDM & Planck15 + LOWZ + CMASS & 0.1426 (12) & 0.298 (14) & 69.2 (17) & \nodata & -1.06 (7) & \nodata \\
$w$CDM & Planck15 + BAO + SN & 0.1424 (11) & 0.307 (9) & 68.1 (10) & \nodata & -1.02 (4) & \nodata \\
\hline
o$w$CDM & Planck15 + LOWZ + CMASS & 0.1425 (14) & 0.297 (21) & 69.4 (26) & +0.0000 (37) & -1.08 (13) & \nodata \\
o$w$CDM & Planck15 + BAO + SN & 0.1424 (13) & 0.308 (9) & 68.0 (10) & +0.0004 (26) & -1.01 (5) & \nodata \\
\hline
$w_0w_a$CDM & Planck15 + LOWZ + CMASS & 0.1427 (13) & 0.370 (36) & 62.4 (32) & \nodata & -0.33 (33) & -1.88 (83) \\
$w_0w_a$CDM & Planck15 + BAO + SN & 0.1429 (13) & 0.311 (10) & 67.8 (10) & \nodata & -0.91 (10) & -0.44 (39) \\
\hline
o$w_0w_a$CDM & Planck15 + LOWZ + CMASS & 0.1422 (14) & 0.364 (36) & 62.8 (32) & -0.0023 (40) & -0.35 (30) & -2.04 (76) \\
o$w_0w_a$CDM & Planck15 + BAO + SN & 0.1423 (14) & 0.313 (10) & 67.5 (11) & -0.0038 (35) & -0.83 (13) & -1.02 (68) \\
\hline
\end{tabular}
\end{table*}

\begin{figure*}
\includegraphics[width=0.4\textwidth]{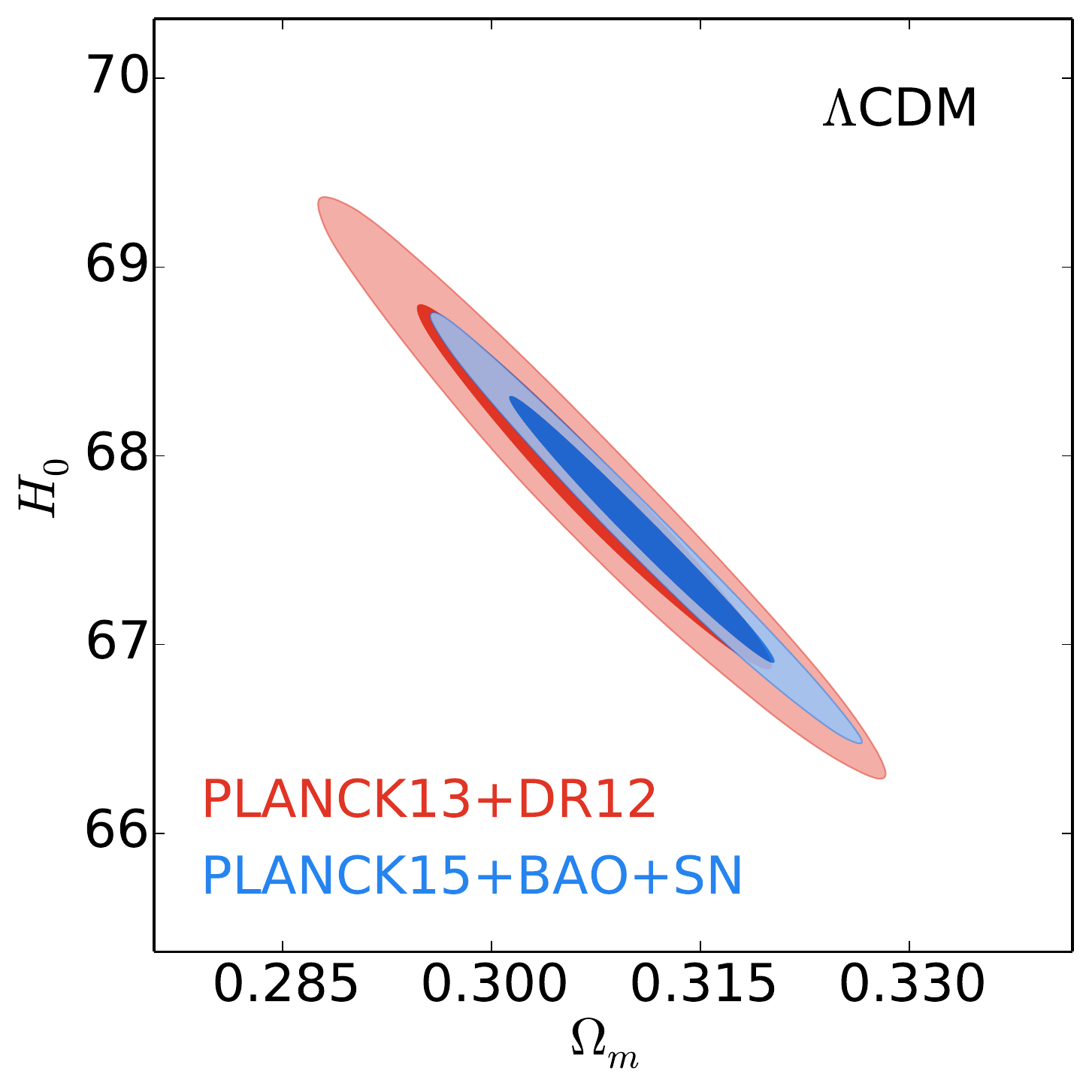}
\includegraphics[width=0.4\textwidth]{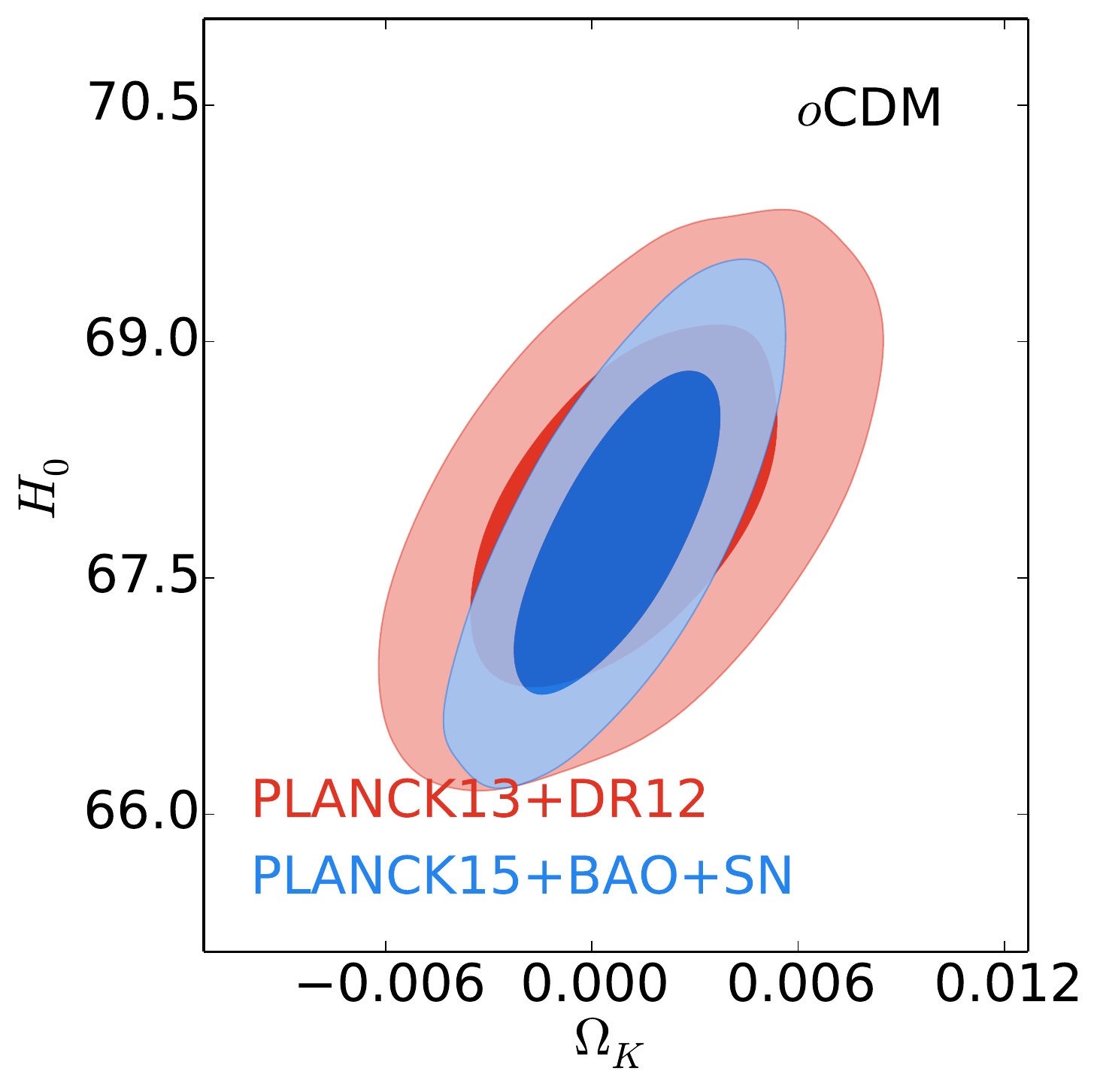}
\\
\includegraphics[width=0.4\textwidth]{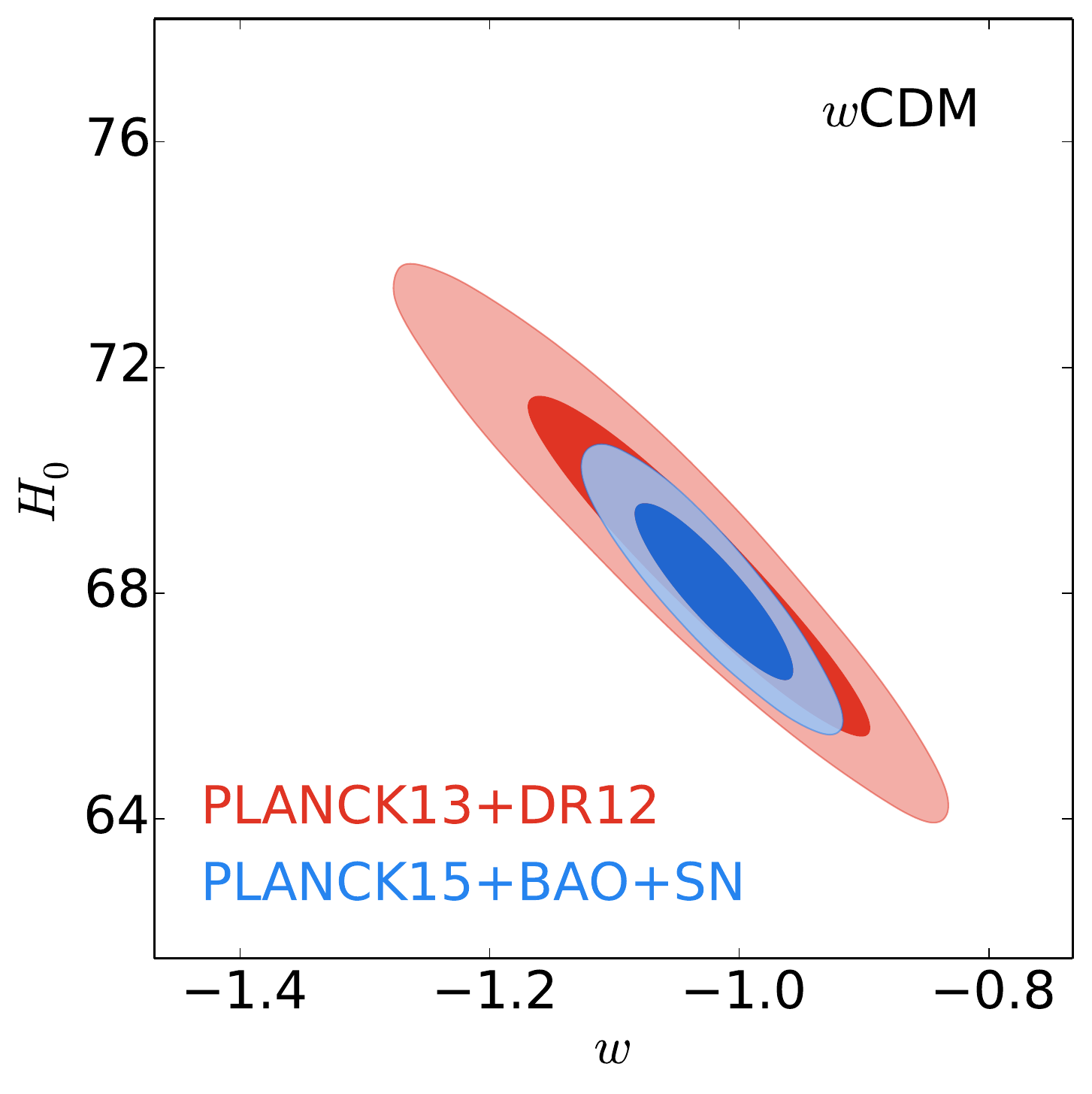}
\includegraphics[width=0.4\textwidth]{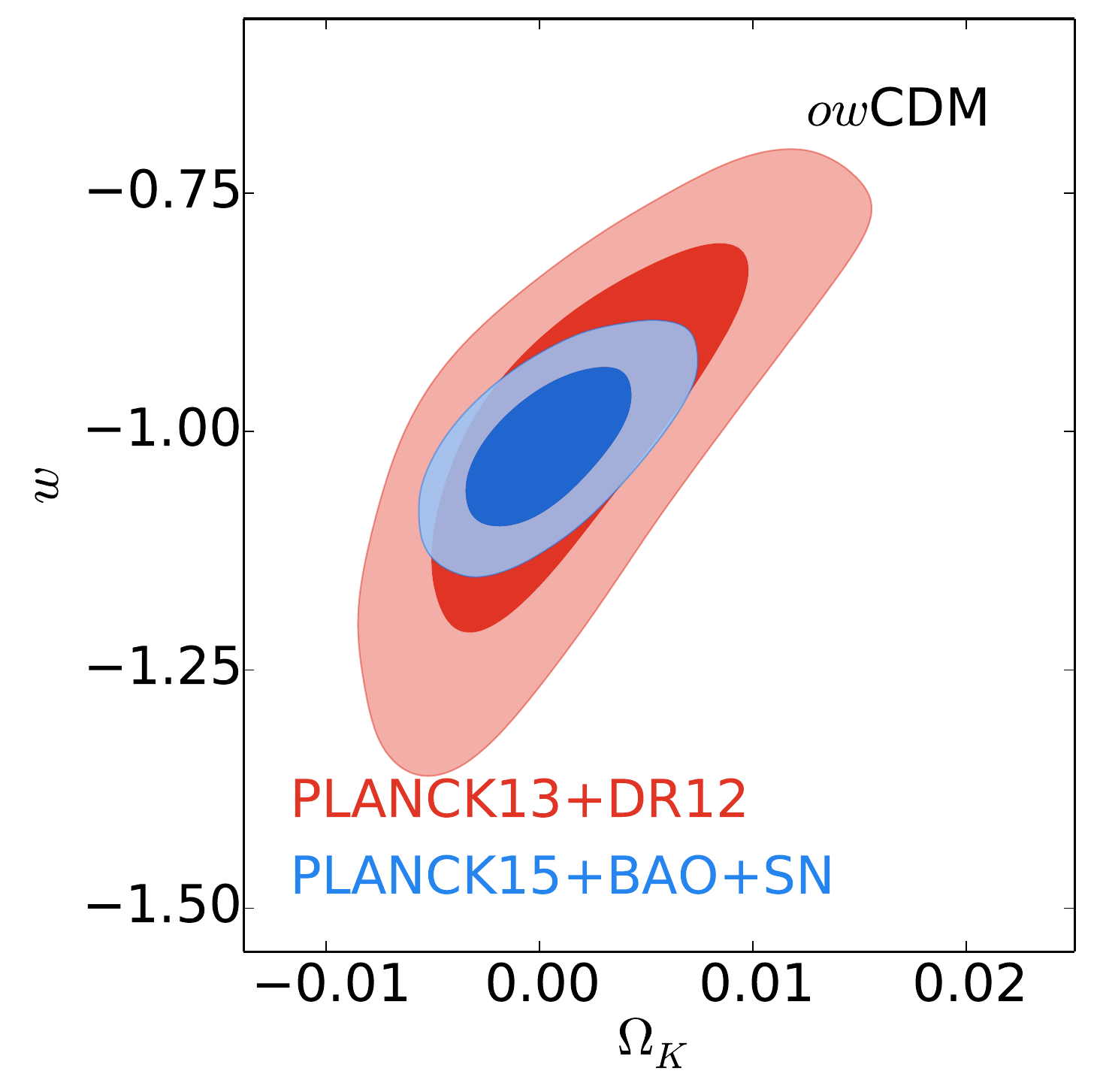}
\\
\includegraphics[width=0.4\textwidth]{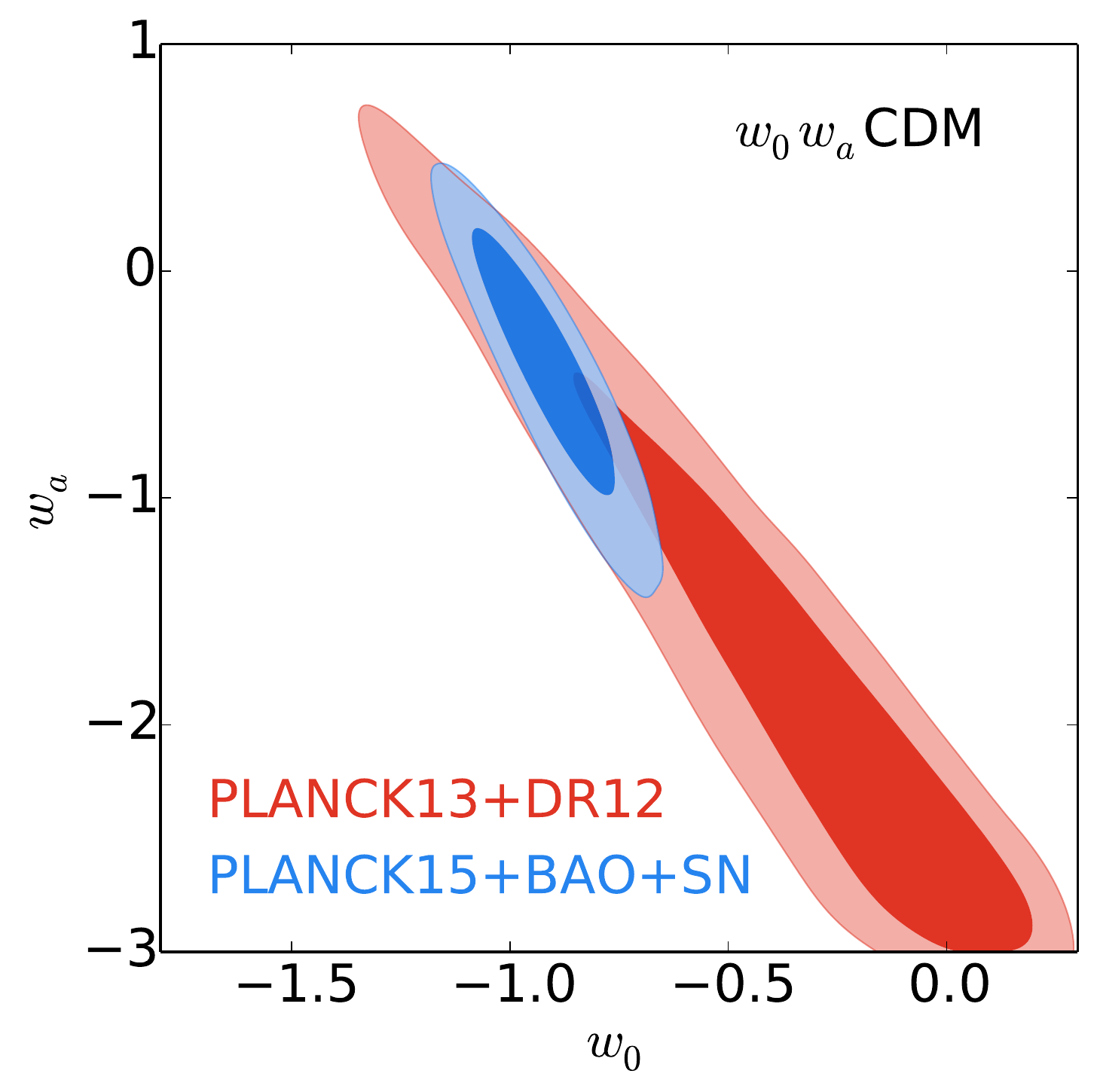}
\includegraphics[width=0.4\textwidth]{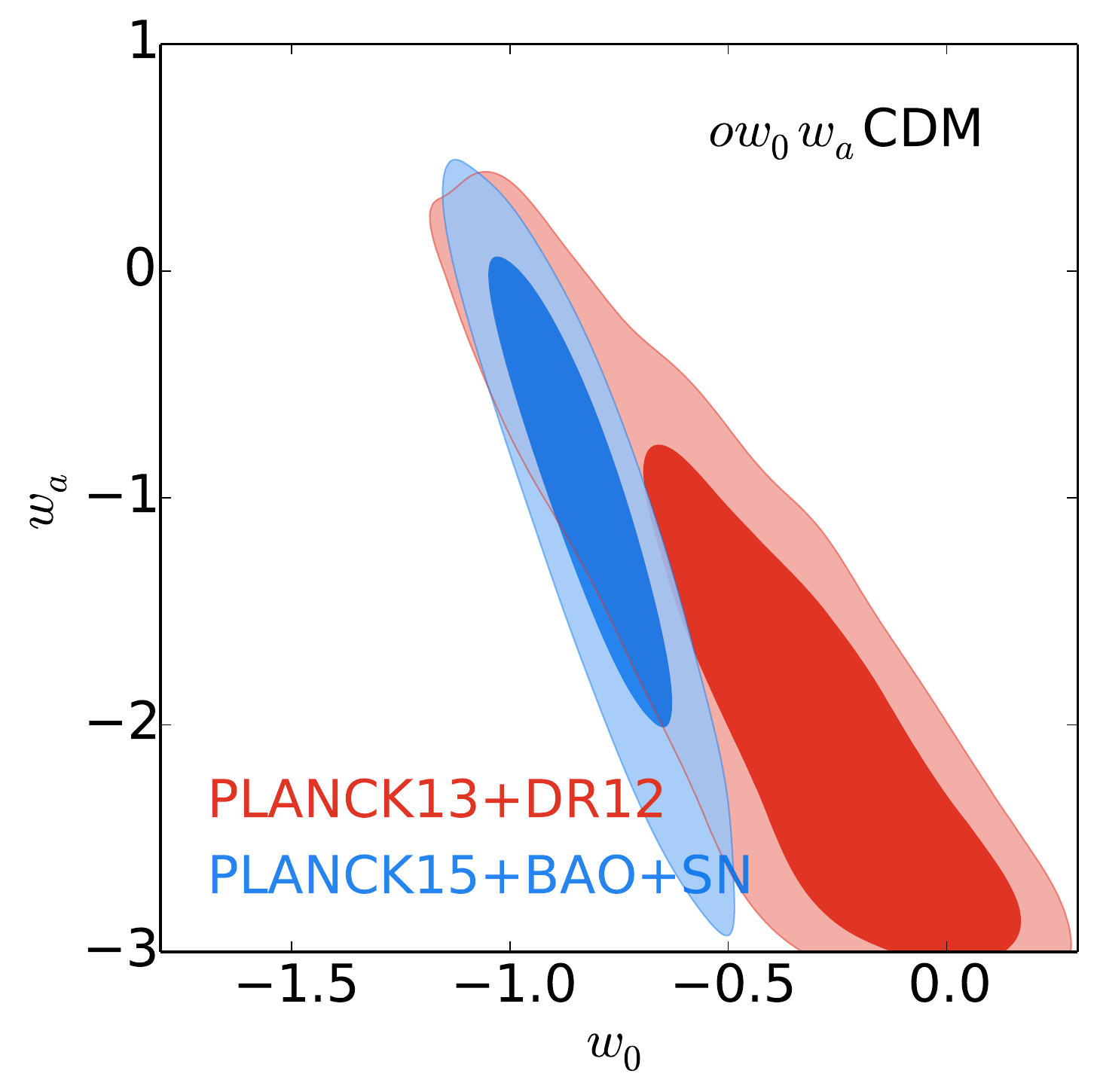}
\caption{Cosmological constraints for different cosmological models, using a combination of Planck15 TT+TE+EE power spectra, JLA SN data, and our DR12 BAO measurements (blue contours). Also shown for comparison are the constraints from Figure~\ref{fig:cosmology} where no supernovae data is used and Planck15 CMB data is replaced with Planck13 (red contours). The left panels assume a flat Universe: $\Lambda$CDM (top panel), $w$CDM (middle panel), and $w_0w_a$CDM (bottom panel). In the right panels the curvature is a free parameter: $o$CDM (top panel), $ow$CDM (middle panel), and $ow_0w_a$CDM (bottom panel). The following priors are assumed: $-0.1<\Omega_k<+0.1$, $-3<w_0<+1$, and $-3<w_a<+3$.}
\label{fig:cosmology2015}
\end{figure*}

Since the results in Table~\ref{tab:planck2015} are more constraining, we will focus on them. The $\Lambda$CDM constraint on the Hubble constant has an error bar nearly half its size for the CMB only case in \cite{Planck2015Cosmo}, creating tension for any reported values of $H_0$ larger than 70 km s$^{-1}$ Mpc$^{-1}$. This result follows the recent analyses of \textit{inverse distance ladder} measurements of \cite{JOINTBAO,Heavens2014,Cuesta2015}. The $\Lambda$CDM cosmology is completely consistent with the fiducial cosmology adopted in \citet[in preparation]{Anderson2015}. The curvature is also reported here with an error bar half its size in \cite{Planck2015Cosmo} for the CMB+BAO+JLA+$H_0$ dataset combination ($\Omega_{\rm K}=0.0008\pm0.0040$) and is consistent with flatness. The equation of state of dark energy is also reported with an error bar half its size in \cite{Planck2015Cosmo} for the CMB+BAO+JLA+$H_0$ dataset combination ($w=-1.02\pm0.08$) and is consistent with a cosmological constant. The figure of merit from the combination Planck13+DR11+Union2 increases by a factor of 1.8 when using Planck15+DR12+JLA (see Figure~\ref{fig:cosmology2015}). 

Finally, if one is interested in combining the updated results from CMASS with the WiggleZ \citep{Blake2011,Kazin2014} results, one can use the correlation matrix in \cite{Beutler2015}, updated to the actual effective volume in CMASS Data Release 12:
\begin{equation}
\tilde{R}=\left(
\begin{array}{cccc}
1     &      &      &   \\
0.039 & 1    &      &   \\
0.020 & 0.57 & 1    &   \\
0.014 & 0.39 & 0.51 & 1
\end{array}
\right)
\end{equation} 
and build the covariance matrix $\tilde{C}=V^{\mathrm{T}}\tilde{R}V$ with $V=(21, 65, 200, 86)$ Mpc. This covariance matrix contains the errors and covariances of the angle-averaged distance measurements from CMASS, the CMASS-WiggleZ cross-correlation, WiggleZ, and the high redshift WiggleZ samples. The corresponding data vector would be $D=(2028,2132,2100,2516)$ Mpc. Although we do not use this in our cosmological constraints, we find it useful since in those cases where the anisotropic constraints from CMASS are not much better than the isotropic ones (see Table~\ref{tab:cosmo}), one can benefit from the extra amount of information from WiggleZ BAO measurements and their cross-correlation with CMASS.

\section{Conclusions}
\label{sec:conclusion}
In this paper we have measured the position of the baryon acoustic peak in the monopole and quadrupole of the correlation function of the CMASS and LOWZ samples from the Data Release 12 of the Baryon Oscillation Spectroscopic Survey (BOSS). The BAO peak has been detected with high significance in both samples, representing the best detection of baryon acoustic oscillations ever done by any galaxy survey so far. Using a large set of QPM mock catalogues that reproduce the clustering of the data catalogues we have tested our fitting methodology, which we find unbiased. From the CMASS sample we have measured the distance scale to $z=0.57$ as $D_V(z)r^{\rm fid}_{\rm d}/r_{\rm d}=2028\pm21$ Mpc, where $r^{\rm fid}_{\rm d}=147.10$ Mpc. The LOWZ sample measures the distance to $z=0.32$ as $1264\pm22$ Mpc. We have also performed the anisotropic fitting of the CMASS correlation function, and for the first time, of the LOWZ correlation function. From this analysis we find an angular diameter distance to $z=0.57$ of $D_{\rm A}(z)r^{\rm fid}_{\rm d}/r_{\rm d}=1401\pm21$ Mpc and a distance to $z=0.32$ of $981\pm20$ Mpc. We also find a Hubble parameter at $z=0.57$ of $H(z)r_{\rm d}/r^{\rm fid}_{\rm d}=100.3\pm3.7$ km s$^{-1}$ Mpc$^{-1}$ and a value at $z=0.32$ of $79.2\pm5.6$ km s$^{-1}$ Mpc$^{-1}$. These values show an excellent agreement with a $\Lambda$CDM model with cosmological parameters given by he recent \textit{Planck} 2015 results.

These constraints on the distance scale to $z=0.32$ and $z=0.57$ are similar to those in Data Release 11 due to the marginal difference in volume, although a small improvement is seen in LOWZ measurements. Nevertheless, these results can be considered more robust than in DR11 thanks to the improved set of mock catalogues using a novel methodology, and also due to an updated systematics weighting scheme. We realize that the predictions on the precision of the cosmic distance scale measurements made at the beginning of the BOSS survey have been partially met. The forecasted measurement precision for angular diameter distance $D_{\rm A}(z)$ was 1.0 per cent, 1.0 per cent, and 1.5 per cent at $z=$0.35, 0.6, and 2.5, respectively, and the forecast precision for the Hubble parameter $H(z)$ was 1.8 per cent, 1.7 per cent, and 1.2 per cent at the same redshifts \citep{BOSSForecasts}. Here we report a 1.5 per cent and 2.0 per cent measurement for $D_{\rm A}(z)$ at $z=$0.57 and 0.32, and a 3.7 per cent and 7.1 per cent measurement for $H(z)$ at the same redshifts, respectively. It remains to be studied whether the discrepancy between the forecasted uncertainties and the measured ones (especially in $H(z)$) is due to a missing ingredient not included in the forecasts (in which case the future is bright for upcoming surveys to reduce the statistical errors), or on the contrary that the measurement is limited by systematics, in which case a larger survey covering our redshift range might not actually help. Whatever the case may be, the values presented in this paper should be considered as an update of those reported in \cite{Anderson2014}, and the final measurements from the BOSS survey will be reported in \citet[in preparation]{Anderson2015}, in which the CMASS and LOWZ samples from Data Release 12 are combined into a single galaxy catalogue that includes additional galaxies that have traditionally been excluded from the LOWZ sample \citep[][companion paper]{Reid2015} but will be recovered in that analysis.

The cosmological constraints reported in this paper have largely benefited from updated cosmological datasets such as the Type-1a Supernovae compilation in \cite{Betoule2014} and the CMB polarization measured by the \textit{Planck} satellite \citep{Planck2015Overview}. We find an improvement of the dark energy Figure of Merit of a factor of 1.8 with respect to \cite{Anderson2014}. Moreover we find that the flat $\Lambda$CDM model is an excellent fitting to the combination of CMB, BAO, and SN datasets. The values we derive for the cosmological parameters include a curvature parameter of $\Omega_k=+0.0008\pm 0.0020$, consistent with a flat geometry of the Universe, and the equation of state of dark energy being $w=-1.02\pm 0.04$, completely consistent with a cosmological constant. In the $\Lambda$CDM model, the Hubble parameter is found to be $H_0=67.6\pm0.4$ km s$^{-1}$ Mpc$^{-1}$, which does not alleviate the tension with direct measurements of the expansion rate, assuming that the sound horizon scale is known with the precision claimed by \textit{Planck} for this cosmological model.

Having reached the milestone of the 1 per cent precision on the distance scale using BAO, which was already achieved by the BOSS survey since Data Release 11, the future of BAO measurements is promising. Improvements are expected in the next few years extending to higher redshifts with the extended BOSS survey \citep{eBOSS} and HETDEX \citep{HETDEX}. Substantial improvements are not expected until results are available from the next generation of experiments, including EUCLID \citep{EuclidAssesment, EuclidDefinition}, LSST \citep{LSST}, SKA \citep{SKA}, WFIRST \citep{WFIRST, WFIRSTReport}, and DESI \citep{DESIOverview, DesiWhitePaper}. It is undeniable, however, that the legacy of BOSS will provide an invaluable guide to analyse and interpret these surveys.

\section{Acknowledgements}
AJC and LV are supported by supported by the European Research Council under the European Community's Seventh Framework Programme FP7-IDEAS-Phys.LSS 240117. Funding for this work was partially provided by the Spanish MINECO under projects AYA2014-58747-P and MDM-2014-0369 of ICCUB (Unidad de Excelencia 'Mar{\'\i}a de Maeztu'). 

Based on observations obtained with Planck (\url{http://www.esa.int/Planck}), an ESA science mission with instruments and contributions directly funded by ESA Member States, NASA, and Canada.

Funding for SDSS-III has been provided by the Alfred P. Sloan Foundation, the Participating Institutions, the National Science Foundation, and the U.S. Department of Energy Office of Science. The SDSS-III web site is \url{http://www.sdss3.org/}.
SDSS-III is managed by the Astrophysical Research Consortium for the Participating Institutions of the SDSS-III Collaboration including the University of Arizona, the Brazilian Participation Group, Brookhaven National Laboratory, Carnegie Mellon University, University of Florida, the French Participation Group, the German Participation Group, Harvard University, the Instituto de Astrofisica de Canarias, the Michigan State/Notre Dame/JINA Participation Group, Johns Hopkins University, Lawrence Berkeley National Laboratory, Max Planck Institute for Astrophysics, Max Planck Institute for Extraterrestrial Physics, New Mexico State University, New York University, Ohio State University, Pennsylvania State University, University of Portsmouth, Princeton University, the Spanish Participation Group, University of Tokyo, University of Utah, Vanderbilt University, University of Virginia, University of Washington, and Yale University.

This research used resources of the National Energy Research Scientific Computing Center, which is supported by the Office of Science of the U.S. Department of Energy under Contract No. DEAC02-05CH11231. Power spectrum computations were supported by the facilities and staff of the UK Sciama High Performance Computing cluster supported by SEPNet and the University of Portsmouth.
Power spectrum calculations, and fitting made use of the COSMOS/Universe supercomputer, a UK-CCC facility supported by HEFCE and STFC in cooperation with CGI/Intel.

The Science, Technology and Facilities Council is acknowledged for support through the Survey Cosmology and Astrophysics consolidated grant, ST/I001204/1.

AJC thanks the hospitality of the Institute of Cosmology \& Gravitation at the University of Portsmouth, where this project was conceived.

\bibliographystyle{mn2e}
\bibliography{baodr12.bib}

\end{document}